%% file: usenix.tex
\documentclass[letterpaper,twocolumn,10pt]{article}
\usepackage{usenix}

\usepackage[ruled,linesnumbered]{algorithm2e}
\usepackage{graphicx}
\usepackage{textcomp}
\usepackage{subfigure}
\usepackage{multirow}
\usepackage{stfloats}
\usepackage{hyperref}
\usepackage{xurl}
\usepackage{diagbox}
\usepackage{caption}
\usepackage{totpages}
\usepackage{bbding}
\usepackage{colortbl}
\usepackage{listings}
\usepackage{pifont}
\usepackage{wasysym}
\usepackage{booktabs}
\usepackage{amssymb}

\usepackage{filecontents}

\begin{document}

\date{}

\newcommand\X{\textsc{VCCL}\xspace}
\newcommand\czt[1]{{\color{red}#1}}
\newcommand\zmh[1]{{\color{blue}#1}}
\newcommand\jym[1]{{\color{green}#1}}
\newcommand\cjh[1]{{\color{cyan}#1}}
\newcommand\zmj[1]{{\color{yellow}#1}}




\title{An Efficient, Reliable and Observable Collective Communication Library in Large-scale GPU Training Clusters}

\author{\large
{\rm Mingjun Zhang\textsuperscript{1}},
{\rm Xiaohe Hu\textsuperscript{1, 3}},
{\rm Menghao Zhang\textsuperscript{2}},
{\rm Ziteng Chen\textsuperscript{1, 4}},
{\rm Yanmin Jia\textsuperscript{1}},
{\rm Yan Zhang\textsuperscript{1}},\\
{\rm Da Liu\textsuperscript{1}},
{\rm Qing Chen\textsuperscript{1}},
{\rm Fangzheng Jiao\textsuperscript{2}},
{\rm Jun Chen\textsuperscript{1}},
{\rm He Liu\textsuperscript{1}},
{\rm Aohan Zeng\textsuperscript{4, 5}},
{\rm Shuaixing Duan\textsuperscript{5}},\\
{\rm Ruya Gu\textsuperscript{1}},
{\rm Yang Jing\textsuperscript{1}},
{\rm Bowen Han\textsuperscript{6}},
{\rm Wei Chen\textsuperscript{1}},
{\rm Wenqi Xie\textsuperscript{1}},
{\rm Jinlong Hou\textsuperscript{3}},
{\rm Yuan Cheng\textsuperscript{3}},\\
{\rm Hongzhou Zhang\textsuperscript{7}},
{\rm Bohua Xu\textsuperscript{6}},
{\rm Mingwei Xu\textsuperscript{4}},
{\rm Chunming Hu\textsuperscript{2}}\\
\textit{\textsuperscript{1}Infrawaves} \quad
\textit{\textsuperscript{2}Beihang University} \quad
\textit{\textsuperscript{3}Shanghai Innovation Institute} \quad
\textit{\textsuperscript{4}Tsinghua University}\\
\textit{\textsuperscript{5}Zhipu AI} \quad
\textit{\textsuperscript{6}China Unicom Research Institute} \quad
\textit{\textsuperscript{7}Shanghai AI Power Technology Co., Ltd}
}


\maketitle

\let\thefootnote\relax\footnotetext{Mingjun Zhang and Xiaohe Hu contributed equally to this paper. Menghao Zhang is the corresponding author (zhangmenghao@buaa.edu.cn).
}

\input{texfiles/0abstract}
\input{texfiles/1introduction}
\input{texfiles/2background}
\input{texfiles/3design}
\input{texfiles/5deployment_2}

\input{texfiles/6operating}
\input{texfiles/7discussion}
\input{texfiles/8relatedwork}
\input{texfiles/9conclusion}

\bibliographystyle{plain}
\bibliography{reference}

\input{texfiles/10appendix}

\end{document}

%% file: texfiles/0abstract.tex
\begin{abstract}

Large-scale LLM training requires collective communication libraries to exchange data among distributed GPUs.
As a company dedicated to building and operating large-scale GPU training clusters, we encounter several practical limitations of NCCL in production, including 1) SM competition between computation and communication, 2) expensive restart costs under link failures, and 3) insufficient observability of transient collective communication anomalies.
To address these challenges, we propose \X, an efficient, reliable, and observable collective communication library in large-scale GPU training clusters.
\X removes SM-consuming P2P kernels by moving intra-node data movement and stream dependency enforcement to CPU threads and GPU copy engines.
\X also introduces a primary-backup QP mechanism to tolerate frequent NIC port failures, and designs a window-based monitor to observe network anomalies at $O(\mu s)$ level.
We open-source \X and deploy it in production training clusters for several months.
Compared with NCCL, \X improves training throughput by up to 5.28\% and reduces massive GPU resource wastage through runtime fault tolerance and fine-grained monitor.
We also share experience and lessons we learned during the deployment of \X in large-scale clusters.

\end{abstract}

%% file: texfiles/1introduction.tex
\section{Introduction}
\label{sec-intro}

Collective communication (CC) plays an essential role for data transmission among GPUs in distributed training of large language models (LLMs).
CC libraries (CCLs) offer various collective primitives and utilize the intra-host and inter-host interconnections, whose communication efficiency is of great importance in large-scale LLM training systems.
Given the dominant market share of NVIDIA GPUs, NVIDIA Collective Communication Library (NCCL) \cite{nccl}, specially tailored and optimized for NVIDIA GPUs, has become the \emph{de-facto} CCL in mainstream large-scale GPU training clusters. 

As a company dedicated to building and operating large-scale LLM training clusters, we strive to provide the most cost-efficient LLM infrastructure for our customers.
However, when dealing with CC for large-scale training clusters, we encounter three main challenges with practical usage of NCCL.

\textbf{Challenge 1: SM competition between computation and communication in P2P primitives.}
GPUs offer high parallelism for general matrix multiplication (GEMM) and value reduction, allowing programmers to write kernels and use GPU streaming multiprocessors (SMs) to accelerate computation.
In NCCL, however, we observe the point-to-point (P2P) primitives (e.g., \texttt{SendRecv} and \texttt{AlltoAll}) that do not involve reductions also consume non-negligible GPU SMs, leading to SM underutilization on computation tasks.
Meanwhile, NCCL P2P suffers from redundant GPU-CPU synchronization irrelevant to the actual communicating process, thereby slowing down the P2P networking performance.
Consequently, NCCL compromises customers' investment of GPU computing power to optimize LLM training efficiency.


\textbf{Challenge 2: Poor tolerance to NIC port failures.}
In large-scale GPU clusters, accidental link failures occur frequently, especially with network speed beyond 400Gbps, leading to stragglers and unexpected crashes of CC due to timeout error.
Nevertheless, NCCL lacks a native fault tolerant mechanism and is therefore unable to migrate the transmission when failures happen.
This can result in significant GPU waste to relaunch CC and LLM training frameworks.

\textbf{Challenge 3: Limited observability for transient network anomalies.}
GPU training clusters are dominated by CC workloads, many of which usually finish within $O(ms)$ and even $O(\mu s)$.
Nonetheless, when network anomalies occur, NCCL lacks fine-grained observability to capture the $O(\mu s)$-level network performance drops at runtime.
It prevents operators from promptly localizing the anomalous links, which is essential to provide proofs for ``network innocence''.

While some existing studies from industrial communities \cite{jiang2024megascale, qian2024alibaba, gangidi2024rdma, an2024fire, hu2024characterization, dong2025evolution, si2025ncclx} have shared their experience in large-scale LLM training, none of them presents a comprehensive and systematic perspective on an ideal CC service with optimized efficiency, reliability, observability in large-scale production environments.
In this paper, we propose \X (Infrawaves Collective Communication Library), an efficient, reliable and observable CCL for large-scale GPU training clusters.
Based on real-world cases and shared requirements from our customers, we present our practices for designing and operating a production-level CCL in detail.
To the best of our knowledge, this is the first work that comprehensively discusses the requirements, designs and operations of a successful CCL in production environment at scale.

\X includes three key designs.
Firstly, \X employs a SM-free P2P mechanism to minimize SM consumption.
Specifically, it utilizes CPU-based alternatives to GPU kernels for intra-node data movement, GPU-CPU synchronization, and CUDA stream dependency management.
SM-free P2P can not only spare more SM resources for GEMMs and improve the P2P performance, but also allow us to optimize the overlaps of pipeline parallelism during LLM training process.
Secondly, \X introduces a primary-backup queue pair (QP) mechanism between each GPU pair.
When a link failure happens, \X switches to another healthy link via backup QP, enabling runtime state migration between QPs and breakpoint retransmission to tolerate NIC port downs.
Thirdly, \X exploits the narrow-waist abstraction of Remote Direct Memory Access (RDMA) to monitor throughput in $O(\mu s)$ granularity.
It records the generation timestamps of Work Requests (WRs) and Work Completions (WCs), and utilizes a sliding window to smooth out fluctuations measured by the naive per-message scheme.

We implement and open-source \X at Github \cite{vccl-github}.
\X has been deployed in multiple large-scale training clusters that we operate, with the largest deployment scaling to 24K GPUs.
Experimental results show that, compared to NCCL, \X delivers higher bandwidth while reducing inter-node small-message latency by 18.9\% on average.
Besides, real-world results show that the SM-free design of \X improves end-to-end training throughput by up to 5.28\% against NCCL.
When RDMA NIC (RNIC) port failures occur, \X's primary-backup QP mechanism can maintain the majority of CC throughput and almost the same training performance as in normal conditions, with production-level data indicating that it can reduce GPU time wastage by nearly 90\%.
For transient network anomalies, \X's window-based monitor accurately pinpoint network stragglers without misclassifying other anomalies in $O(\mu s)$ granularity with low system overhead.
Furthermore, we also share several experiences and lessons when integrating \X with large-scale GPU clusters, hoping this work can give communities more insights on the production-level CCL in LLM training.

The main contributions of this work include:

\begin{itemize}
    \item We identify that current CCLs fall short of realizing efficient, reliable collective communication with sufficient  observability in large-scale GPU clusters (\S\ref{sec-motivation}).
    \item We design \X, comprising a SM-free mechanism for P2P primitives, a primary-backup QP mechanism to tolerate NIC port downs, and a fine-grained monitor for inter-node communication (\S\ref{sec-design}).
    \item We implement and deploy \X, with real-world results validating its efficiency, reliability and observability (\S\ref{sec-deployment}).
    We also share experience and lessons we learned from deploying \X in real-world production GPU clusters (\S\ref{sec-operating}).
\end{itemize}

This work does not raise any ethical issues.

%% file: texfiles/2background.tex
\section{Background and Motivation}
\label{sec-motivation}

\subsection{Preliminary}
\label{subsec-preliminary}

\noindent\textbf{GPU Architecture and Programming Model.} 
Modern GPUs serve as the fundamental accelerator for LLM, primarily comprising multiple SMs and High Bandwidth Memory (HBM). 
Each SM consists of numerous CUDA cores, registers, and shared memory to execute parallel threads. 
Developers leverage the CUDA toolkit \cite{cuda} to submit computational tasks as kernels, which are scheduled onto SMs in the form of thread blocks. 
Beyond GEMM, SMs are also responsible for managing data movement and synchronization logic in communication libraries, making SM efficiency critical for both computation and coordination.

\noindent\textbf{Collective communication library and RDMA.} 
To enable high-performance data exchange between GPUs, NCCL \cite{nccl} is widely adopted. 
NCCL detects the hardware topology (e.g., NVLink, PCIe) and establishes optimized logical channels, such as rings or trees. 
For inter-node communication, NCCL typically relies on RDMA \cite{rdmaverbs} to bypass the CPU kernel and minimize latency. 
RDMA utilizes QP to allow direct memory access between remote GPUs. 
By posting WRs and polling Completion Queues (CQ), RDMA provides the high-bandwidth, low-latency transport layer necessary for large-scale clusters.

\noindent\textbf{Distributed LLM Training.} 
Current LLM training frameworks \cite{pytorch, megatronlm, deepspeed} support multiple parallelism strategies to enable efficient training among distributed GPUs, including:
1) Data Parallelism (DP) uses \texttt{AllReduce} to synchronize gradients; 
2) Tensor Parallelism (TP) and Pipeline Parallelism (PP) partition model weights and layers, relying on \texttt{AllGather} or \texttt{SendRecv} for activations; 
and 3) Mixture-of-Experts (MoE) utilizes \texttt{AlltoAll} for token routing. 
The performance of these distributed strategies is fundamentally constrained by the underlying efficiency of SM-based coordination and RDMA-based network transmission.
With increasing scale and complexity of LLM training, CCL has become more crucial in both academia and industry \cite{nccl, rccl, msccl, deepep, shah2023taccl, wu2024mccs, liu2024rethinking}.

\subsection{Challenges in Large-Scale Training}
\label{subsec-challenge}

Although NCCL has become the \emph{de-facto} CCL in production-level GPU training clusters, we still encounter several challenges in real-world scenarios, falling short of several essential properties of efficiency, reliability and observability.

\begin{figure}[t]
  \centering
  \subfigure[Inter-node P2P.]{\includegraphics[width=0.43\textwidth]{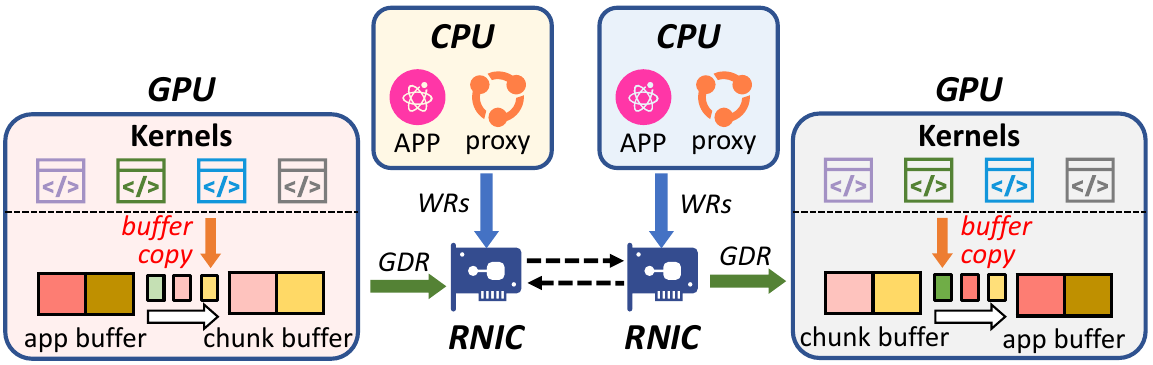}}
  \subfigure[Intra-node P2P.]{\includegraphics[width=0.43\textwidth]{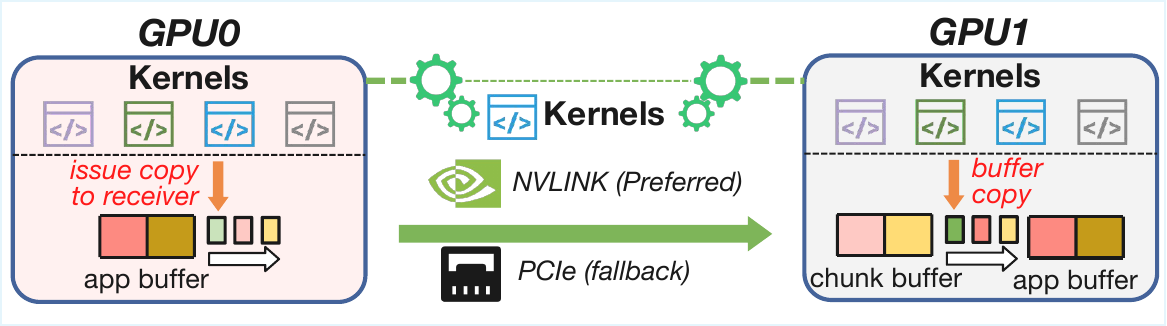}}
  \vspace{-1\baselineskip}
  \caption[Caption for LOF]{NCCL P2P w/ kernels.}
  \label{fig-nccl-p2p}
  \vspace{-1\baselineskip}
\end{figure}

\noindent\textbf{SM competition between computation and communication.}
GPUs utilize SM to execute parallel GEMMs in deep learning, but we observe that CC primitives that do not involve reduction, such as \texttt{SendRecv} used in PP, also consume non-negligible SM resources in NCCL.
This phenomenon significantly hinders the effective overlapping of computation and communication.
Figure \ref{fig-nccl-p2p} illustrates the basic procedure of inter-node and intra-node \texttt{send/recv} primitives of NCCL.
For inter-node P2P operation, NCCL consists of three steps:
(1) Sender buffer copy: sender rank launches kernels to copy data from application buffer to sender buffer, and notify the NCCL proxy thread upon chunk availability.
(2) Network transmission: NCCL proxy thread posts WRs to trigger NIC doorbell, then sender RNIC transmits packets to the remote RNIC by GPUDirect RDMA (GDR) \cite{shi2014designing}.
(3) Receiver buffer copy: receiver rank launches kernels to continuously move data from the receiver buffer to the application buffer as chunks arrive.
On the other hand, for intra-node P2P operation, the procedure follows a similar yet slightly different two-step pattern, and the data movement is driven entirely by kernel launches:
(1) Intra-node transmission: sender rank issues memory copy operation from sender buffer to receiver chunk buffer via NVLink/PCIe.
(2) Receiver buffer copy.


We conduct a NCCL-Tests \cite{nccltest} experiment to evaluate the GPU SM utilization of \texttt{SendRecv} and \texttt{AlltoAll} workloads on 2 servers, each of which is equipped with 8 NVIDIA Hopper GPUs and 8 NVIDIA ConnectX-7 RNICs.
The inter-host topology follows a 1:1 oversubscribed two-tier CLOS topology with 400Gbps links.
And we use Nsight \cite{nsight} and NVIDIA Data Center GPU Manager (DCGM) \cite{dcgmi} to monitor the GPU SM utilization.
Experimental data in Appendix \ref{appendix-nccl-sm-utilization} shows that though \texttt{SendRecv} and \texttt{AlltoAll} are reduction-free operations, NCCL still consumes non-negligible SM resources.
What's worse, SM occupation by kernels will not be freed until the entire P2P process finishes.
For better training efficiency, our customers expect more GPU resources to be reserved for GEMM and to minimize SM wastes in communication tasks.

\begin{figure}[h]
    \includegraphics[width=0.99\linewidth]{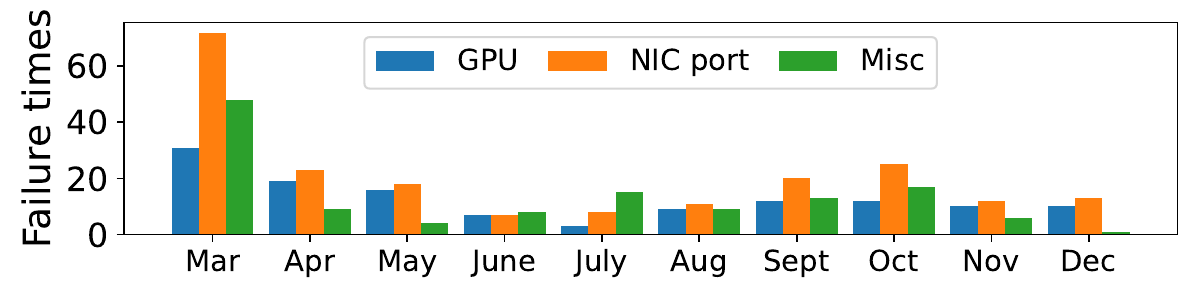}
    \centering
    \vspace{-1\baselineskip}
    \caption{Failure statistics in 2025.}
    \label{fig-down-statistic}
    \vspace{-1\baselineskip}
\end{figure}

\noindent\textbf{Expensive restart costs in link failures.}
As inter-host communication is embracing much higher bandwidth than before, we encounter more frequent downs of RNIC ports within and beyond 400Gbps network bandwidth.
Figure \ref{fig-down-statistic} collects statistics on the failure types over 10 months in 2025 within a GPU cluster.
It is obvious that the link failures, caused by optical modules and RNIC hardware, contribute the most failures than GPUs and miscellaneous types.

In LLM training, link failures can lead to stragglers and suspension of the CC operation, and even crash of the entire training process by timeout exceptions. 
Despite the severe impact, mitigating link failures is typically straightforward, involving simple actions such as reseating optical modules or resetting RNIC driver.
Unlike GPU failures, link failures do not require node isolation, theoretically allowing the original communication groups to remain intact.
Nevertheless, NCCL lacks a built-in mechanism to tolerate the potential downs of RNIC ports, which introduces much time \& financial costs and GPU wastes to restart the training.
Therefore, our customers are seeking a lightweight fault tolerance mechanism to resist frequent RNIC port downs.

\noindent\textbf{Lack of transient network observability.}
In LLM training, problems can occur randomly at every point. 
We observe that existing issues in production clusters can be categorized into two types:
(1) GPU/Network failures: severe GPU/Link failures directly leading to training hang or crash.
(2) GPU/Network stragglers: GPU/Network performance degradation that slows down the training process without completely stopping it.
While existing diagnostics \cite{dcgmi, SNMP} and previous works from industry \cite{deng2025mycroft, dong2025evolution} effectively identify the root causes of persistent faults, they often fail to characterize transient performance degradation.
Consequently, without fine-grained monitoring, network teams cannot rule out network issues as the cause of these fluctuations.

However, current network monitor systems in production clusters are based on hardware counters in $O(s)$ granularity, but most CC operations usually finish within $O(ms)$ and even $O(\mu s)$.
Although NCCL Profiler \cite{ncclprofiler} and Inspector \cite{nccl_inspector} can provide observability for each CC, they lack visibility into the online network performance since several CCs may flow through one NIC simultaneously, so that the tracing of single CC is insufficient to reflect network status.
Therefore, our customers are demanding an observable CCL to monitor transient network anomalies.

\subsection{Existing Works Fall Short}
\label{subsec-existing-work}

Some works have been proposed to optimize CC in different aspects.
For efficiency, Centauri \cite{chen2024centauri}, CoCoNet \cite{jangda2022breaking} and Wang et al. \cite{wang2022overlap} fuse kernels for fewer launches, and HFReduce \cite{an2024fire} offloads the reduction operation of \texttt{allreduce} from GPUs to CPUs, but none of them consider GPU SM competition between computation and communication.
Recently NCCLX \cite{si2025ncclx} introduces a similar SM-free design like \X, nevertheless, it still launches a kernel that occupies a small number of SMs to guarantee the sequential execution of stream computation and communication, which hinders training efficiency (\S\ref{subsec-eval-efficiency}).
Additionally, NCCLX necessitates intrusive modifications to the PyTorch \cite{pytorch} framework, further compromising its deployability and overall usability.

For reliability, industrial practices \cite{jiang2024megascale, gangidi2024rdma, qian2024alibaba, dong2025evolution, an2024fire} tolerate link failures by longer NCCL timeout, traffic engineering, dual-plane switches and faster checkpoints, but they either lack timely analysis \& tolerance on failures, or require hardware supports that hurt deployability.
NCCL itself introduces a dynamic communicator adjustment feature \cite{nccl2.29} in version 2.29.
However, simple link failures do not necessarily require isolating nodes, allowing the original communication group to be maintained.

For observability, early works by Microsoft \cite{guo2016rdma, bai2023empowering} and ByteDance \cite{liu2023hostping, liu2024r} visualize performance anomaly with end-host probing and logging.
However, these works require frequent tracing of QP runtime states, resulting in non-negligible overhead and prolonged profiling times.
More recent works from industry like Mycroft \cite{deng2025mycroft} and Aegis \cite{dong2025evolution} adopt a dependency-based tracing system to track network issues.
However, they struggle to capture transient network dynamics.
NCCL Profiler \cite{ncclprofiler} and Inspector \cite{nccl_inspector} can provide observability for each CC.
However, both fail to reflect real-time network status, as multiple CCs may share one NIC simultaneously.
Although sketches implemented in programmable switches \cite{zheng2024mumon} can monitor $O(\mu s)$ network dynamics, programmable switches are difficult to deploy in production-level clusters.


%% file: texfiles/3design.tex
\section{\X Design}
\label{sec-design}

\subsection{Overview}
\label{subsec-overview}

\begin{figure}[t]
    \includegraphics[width=0.9\linewidth]{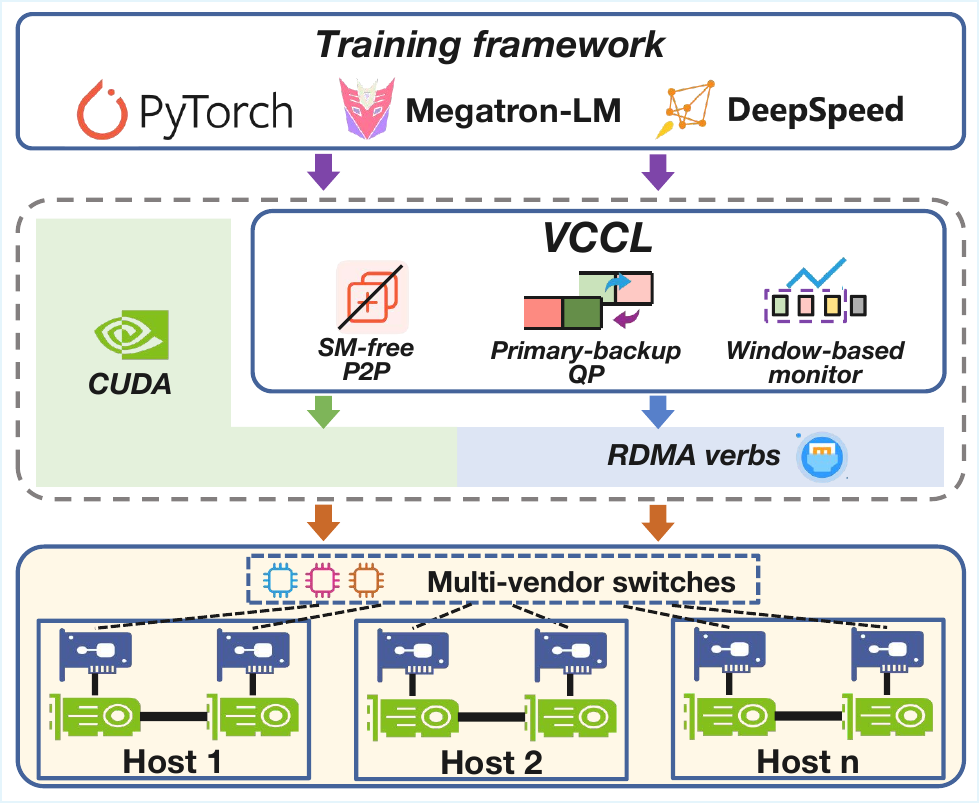
    }
    \centering
    \vspace{-1\baselineskip}
    \caption{\X overview.}
    \label{fig-overview}
    \vspace{-1.5\baselineskip}
\end{figure}

Figure \ref{fig-overview} demonstrates \X's overview.
Generally, \X works as a middleware between upper-layer LLM training frameworks \cite{pytorch, megatronlm, deepspeed} and underlying hardware.
\X uses the unified programming interfaces (CUDA and RDMA verbs) to enable CC among distributed GPUs during the training process.
\X is developed based on NCCL, enabling the communication group with bootstrapping, topology search \& graph construction and channel establishment.
Additionally, \X introduces three extra modules: 
1) SM-free P2P for resource-efficient and performant communication;
2) a primary-backup QP mechanism for fault tolerance.
3) a window-based monitor to capture network status and potential anomalies.
These three modules collaborate to achieve an efficient, reliable and observable CCL.

\subsection{SM-free P2P}
\label{subsec-kernelfree}

The contention for SMs between computation and communication (\S\ref{subsec-challenge}) constrains the potential for training efficiency. 
In contrast, CPUs typically exhibit lower utilization and greater resource availability compared to GPUs \cite{jiang2020unified, hu2024characterization, an2024fire}.
Consequently, offloading P2P operations from GPUs to CPUs seems a viable strategy.
By eliminating GPU kernel launches, we can reclaim SM resources for general computation tasks, thereby accelerating the training process.

\noindent\textbf{SM-free P2P in \X. }
The function of launched kernels in NCCL can be categorized into three aspects:
(1) Intra-node data movement: Conducting intra-node communication or memory copies between application buffers and chunk buffers.
(2) GPU-CPU synchronization: Coordinating the CPU-based control plane (Proxy) and the GPU-based data plane via shared pointer.
(3) Dependency management on CUDA stre\allowbreak ams: Enforcing strict execution ordering among different streams. 
Given that training frameworks often pre-dispatch operations to enable asynchronous communication, inserting post-kernel events ensures that communication only starts after prerequisite computations finish and blocks subsequent dependent tasks.
Here we detail how our SM-free P2P design effectively realizes the functionalities discussed above without relying on GPU kernels.

\noindent \underline{1) Intra-node data movement. } 
\begin{figure}[t]
  \centering
  \subfigure[caption for LOF][Inter-node SM-free P2P. \footnotemark]{
    \includegraphics[width=0.94\linewidth]{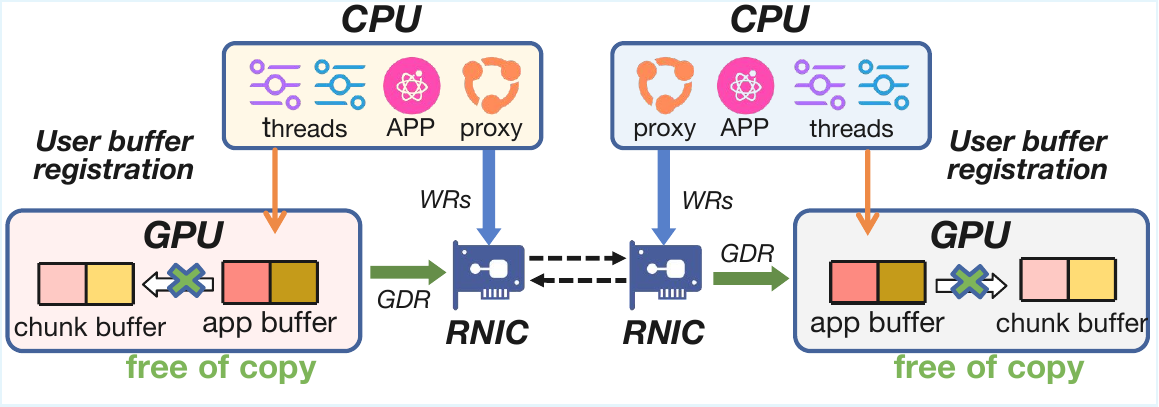}
    \label{fig-smfree-inter}}
  \subfigure[Intra-node SM-free P2P.]{
    \includegraphics[width=0.94\linewidth]{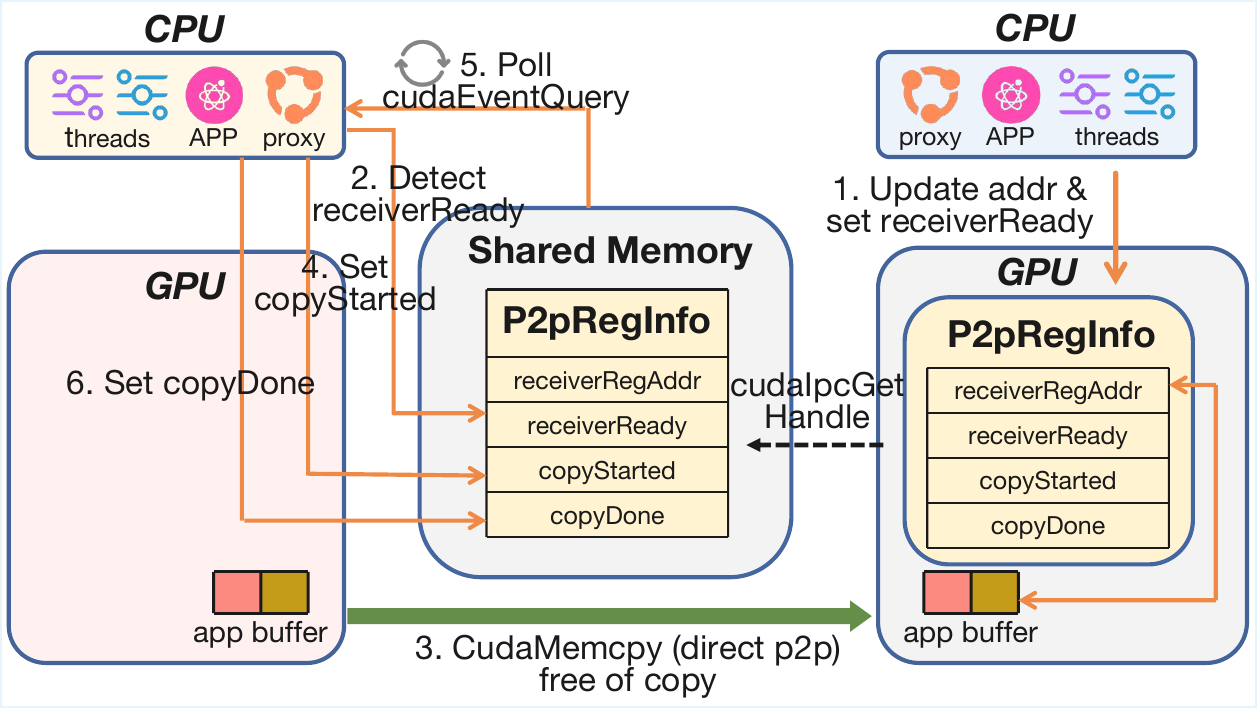}
    \label{fig-smfree-intra}}
  \vspace{-1\baselineskip}
  \caption{SM-free P2P in \X.}
  \label{fig-vccl-design}
  \vspace{-1.0\baselineskip}
\end{figure}
\footnotetext{Due to page limit, we only draw inter-node SM-free P2P between ranks sharing the same local index.}
For P2P communication between ranks sharing the same local index or disable \texttt{PXN} \cite{nccl_pxn}, as depicted in Figure \ref{fig-smfree-inter}, where application and chunk buffers reside on the same GPU, \X leverages a zero-copy design inspired by NCCL's User Buffer Registration feature \cite{nccl_registered_buffer}. 
When an upper-layer application initiates a \texttt{SendRecv} operation, \X first intercepts the memory registration request from the training framework. 
It then registers a user buffer via \texttt{ncclMemAlloc} and returns a handle, thereby enabling direct P2P data transmission from the application buffer and bypassing intermediate copies to the chunk buffer.
Conversely, if \texttt{PXN} enabled, for communication between ranks with different local index, intra-node memory copies remain essential, since directly fetching data from application buffers by GDR incurs prohibitive PCIe overhead and potential cross-NUMA traversal.
Therefore, we utilize \texttt{cudaMemcpy} on the sender side for data movement, relying on the GPU's copy engines instead of launched kernels.
The CPU Proxy then subsequently polls for completion via \texttt{cudaEventQuery} and initiates remote transmission upon \texttt{cudaSuccess}.

For intra-node P2P communication, as shown in Figure \ref{fig-smfree-intra}, we leverage a shared memory design.
The receiver maintains a shared buffer, \texttt{P2pRegInfo}, containing specific four fields.
Three of them---\texttt{recei\allowbreak ver\allowbreak Ready} (trigger data movement),  \texttt{copy\allowbreak Started} (confirm \texttt{cudaMemcpy} issuance), and \texttt{copyDone} (track operation completion)---are 1-bit flags to coordinate the sender and receiver.
The fourth field, \texttt{receiver\allowbreak RegAddr}, is a dynamic pointer, pointing to the receiver's application buffer to enable zero-copy transfer.
During Connection phase, we utilize \texttt{cudaIpcGetHandle} to map this buffer into a shared region.
During communication phase, the receiver first updates 
\texttt{receiver\allowbreak RegAddr} and changes \texttt{receiverReady} to 1 to notify the sender.
Upon detection, the sender issues \texttt{cudaMemcpy}, sets \texttt{copyStarted} and subsequently polls for completion via \texttt{cudaEventQuery}. 
Finally, the sender sets \texttt{copyDone} to finalize the transmission.

\noindent \underline{2) GPU-CPU synchronization. } 
In standard NCCL, coordination relies on GPU-CPU synchronization via shared volatile variables (e.g., $\textit{posted}$). 
The CPU proxy must continuously poll these flags to detect the completion of kernel-side data movement before issuing subsequent operations, such as posting WRs. 
This polling mechanism inevitably incurs significant synchronization latency.
By migrating data-plane control from GPU to CPU, \X effectively eliminates the overhead associated with GPU-CPU synchronization. 
For inter-node P2P, we utilize CPU-based memory copy operations to facilitate pipelined transmission. 
Similarly, for intra-node P2P, the aforementioned shared memory design is employed for state synchronization, ensuring that the process remains entirely free of GPU-CPU synchronization costs.

\begin{figure}[t]
    \includegraphics[width=0.94\linewidth]{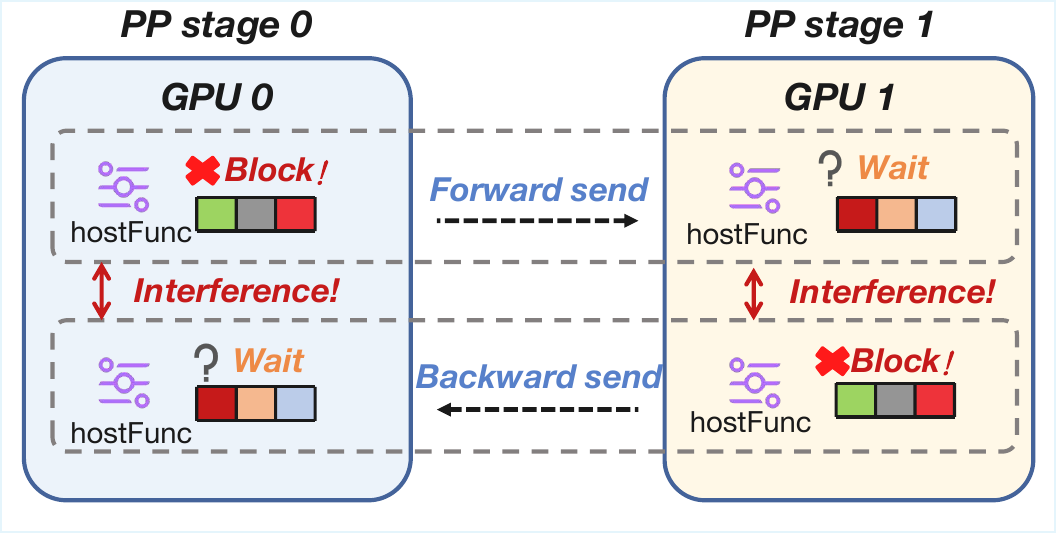}
    \centering
    \vspace{-1\baselineskip}
    \caption[Caption for LOF]{Potential hang with hostFunc.}
    \label{fig-hostfunc-limitation}
    \vspace{-1.5\baselineskip}
\end{figure}

\noindent \underline{3) Dependency management on CUDA streams. } 
To address the absence of kernels on the CUDA stream, we first utilize \texttt{cuda\allowbreak Launch\allowbreak HostFunc} to manage stream execution order via the CPU.
Specifically, \X replaces standard kernels with host functions to maintain dependency constraints.
\X employs two synchronization primitives: \texttt{proxyReadyEvent} (to signal buffer readiness) and \texttt{proxyOpCount} (to track pending operations).
The host function first awaits prerequisite computations and then asserts \texttt{proxyReadyEvent} to enable subsequent communication. 
It then polls \texttt{proxyOpCount} until it reaches zero, effectively blocking dependent tasks.
However, we observe occasional hang during training with this SM-free design. 
After deep investigation, we find this issue stems from the single-thread limitation of \texttt{hostFunc}.
As illustrated in Figure \ref{fig-hostfunc-limitation}, a deadlock arises when the host function on GPU 0 blocks waiting for the \texttt{proxyReadyEvent} of backward communication from GPU 1. 
Simultaneously, GPU 1 awaits forward communication from GPU 0; however, GPU 0 cannot trigger the necessary ready signal because its host execution thread is blocked. 
To resolve this, we initially modify the training framework (e.g., Megatron \cite{megatronlm}) to merge bidirectional P2P communication groups (GPU 0 $\leftrightarrow$ GPU 1), successfully eliminating the hang.

However, this approach compromises deployability, as it requires modifications to frameworks.
We explore a fully non-intrusive alternative that preserves framework compatibility without requiring any code changes.
We point out a key limitation of \texttt{cudaLaunchHostFunc} is that it is not a stream-native progress mechanism: it invokes CPU callbacks through CUDA's internal host execution path, where callbacks from independent streams may be serialized.
In contrast, \texttt{cuStreamWriteValue} and \texttt{cuStreamWaitValue} are stre\allowbreak am memory operations enqueued directly into CUDA streams.
They update or wait on memory locations in stream order without relying on host-callback execution, thereby avoiding serialization-induced deadlocks while preserving dependency constraints.
By adopting the two stream memory operations, we devise a mechanism where \texttt{cuStream\allowbreak Write\allowbreak Value} updates the \texttt{proxy\allowbreak Ready\allowbreak Event} and \texttt{cuStream\allowbreak WaitValue} blocks the CUDA stre\allowbreak am until \texttt{proxy\allowbreak OpCount} returns to zero. 
This approach guarantees strict execution ordering without modifying the training framework.

\noindent\textbf{Improve overlaps in PP.}
Pipeline parallelism \cite{megatronlm} uses P2P to synchronize the activations/gradients of forward/backward passes among GPU workers.
However, P2P communication in NCCL has to compete for GPU SM resources with forward \& backward passes, leading to compromised computation performance during overlap.
In contrast, \X offloads P2P communication from GPUs to CPUs, and thus spares the SM resources for computation, empowering faster forward \& backward passes and better overlaps to hide the P2P communication overhead.

\begin{figure}[t]
    \includegraphics[width=1\linewidth]{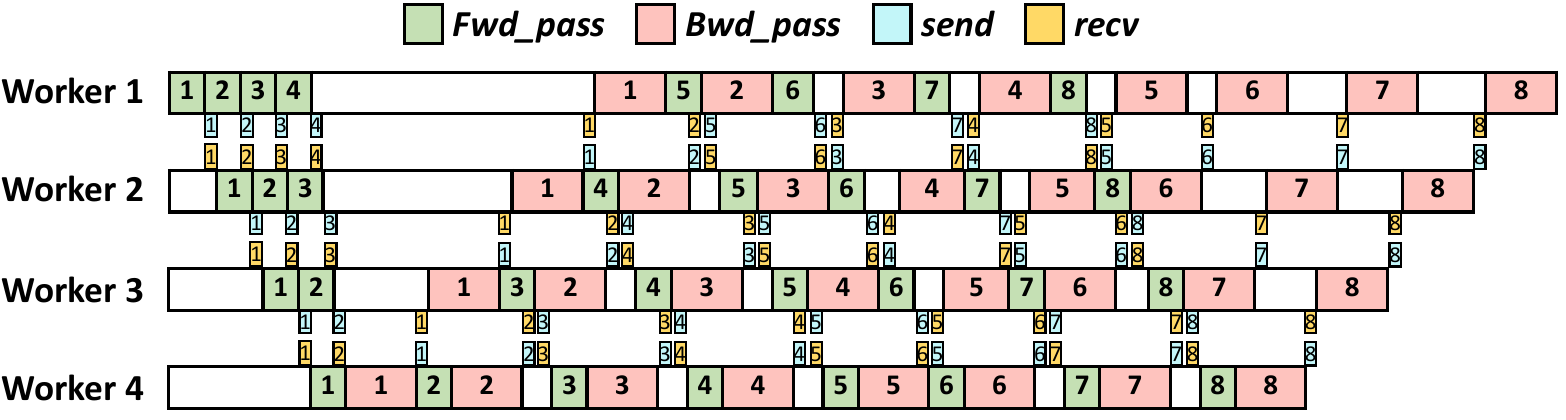}
    \centering
    \vspace{-1.5\baselineskip}
    \caption[Caption for LOF]{PP with \X.\footnotemark}
    \label{fig-pp}
    \vspace{-1.5\baselineskip}
\end{figure}

Figure \ref{fig-pp} shows \X's PP progress using 1F1B strategy \cite{narayanan2019pipedream} with 8 microbatches and 4 GPU workers.
Unlike kernel-based P2P that conducts computation and P2P communication in sequence, \X overlaps the P2P communication with forward \& backward passes during the PP process.
Generally, after finishing a forward/backward pass of a microbatch, \X allows the worker to proceed to the backward/forward pass of the next microbatch, and simultaneously \texttt{send} the activation/gradient to the next/previous worker.
As a receiver, the worker can also \texttt{recv} the results from a sender, while performing a forward or backward pass at the same time.
Compared to naive PP by NCCL, \X overlaps the P2P communication and enables faster forward \& backward passes with more SM resources, therefore accelerating the entire LLM training.

\subsection{Fault Tolerance with Backup QPs}
\label{subsec-fault-tolerance}



\X adopts a primary-backup QP mechanism to tolerate link failures.
Though previous works like Mooncake\cite{qin2025mooncake} propose a similar fail-safe solution, to the best of our knowledge, \X is the first work that successfully integrates such a feature into CCL with production validation. 
We mainly detail the underlying designs in this subsection.

\noindent\textbf{Creation of Backup QPs.}
During the bootstrap for the inter-node connection, in addition to a primary QP that uses the closest RNIC, \X creates a backup QP by using the second-closest RNIC for each GPU.
If a link fails due to port downs, \X migrates the traffic from the primary QP to the backup QP, and retransmits the data at the breakpoint.
\X also monitors the failed link, and when it is fixed, \X turns back to the primary QP for faster transmission.
Note that if the RNIC has dual ports, \X creates the backup QP on the other port of the same RNIC, enjoying the same hardware distance as the primary QP.

\begin{figure}[t]
  \centering
  \subfigure[Case 1: triggered by local WC.]{\includegraphics[width=0.43\textwidth]{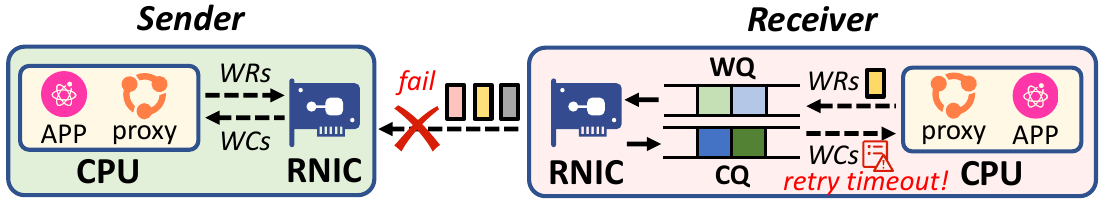}}
  \vspace{-0.5\baselineskip}
  \subfigure[Case 2: triggered by timeout threshold.]{\includegraphics[width=0.43\textwidth]{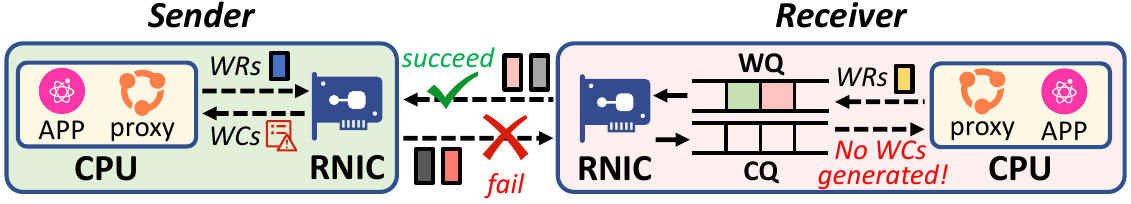}}
  \vspace{-0.5\baselineskip}
  \caption{Perception of link failures.}
  \label{fig-backup-trigger}
  \vspace{-1.5\baselineskip}
\end{figure}

\noindent\textbf{Failure perception.}
When an RNIC port down occurs, the sender cannot determine whether the inflight data has been received by the receiver. 
To address this, we employ a receiver-driven QP switching strategy, leveraging NCCL's built-in \texttt{CTS} (clear to send) message mechanism to detect link failures.
We define two trigger conditions for QP switching to handle different failure scenarios.
Figure \ref{fig-backup-trigger}(a) illustrates the first case, where a receiver-to-sender communication failure prevents the receiver from notifying the sender via \texttt{CTS}.
In this case, after exhausting a pre-defined number of retries and exceeding \X's retry timeout threshold (determined by \texttt{ICCL\_IB\_TIMEOUT} and \texttt{ICCL\_IB\_RETRY\_CNT}), the receiver RNIC generates a WC error.
This error notifies the Proxy of the communication failure, allowing the control plane to detect the port-down event and trigger QP switching.
Notably, we intentionally retain this retry window because many failures in our deployments are short-lived link flaps, about half of which recover within seconds. 

The second case, illustrated in Figure \ref{fig-backup-trigger}(b), presents a more complex scenario: the RNIC port fails after the receiver successfully transmits a \texttt{CTS} message, but before the subsequent data transmission from the sender completes.
In this case, while the sender eventually receives a Work Completion (WC) retry timeout error, the receiver remains unaware of this failure and generates no local WC error to alert the control plane.
To address this, we design a timeout-based trigger mechanism.
When a WR is issued, the receiver records its timestamp and monitors the corresponding WC.
If no WC is generated within the period $\delta$, the receiver proactively resends a \texttt{CTS} message to probe the link state. If this probe fails, the receiver generates a local WC error, triggering the QP switching process as described in the first case.
This ``double-check'' mechanism is critical; it prevents the system from misclassifying a healthy link as faulty when a sender is simply stalled due to upstream data dependencies---a common occurrence in collective communication processes. 
We set $\delta$ slightly larger than the retry-timeout threshold to account for queuing and propagation delays.

\noindent\textbf{State synchronization and migration.}
\begin{figure}[t]
    \includegraphics[width=0.9\linewidth]{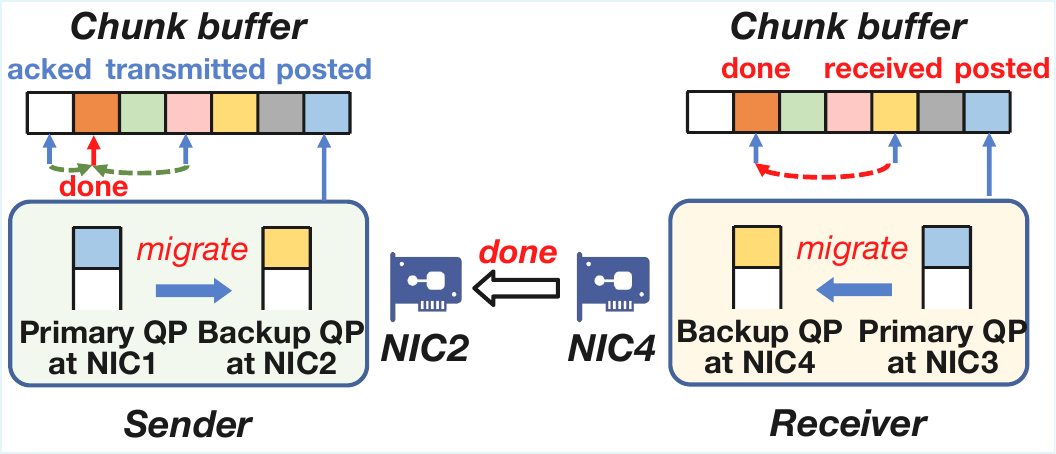}
    \centering
    \vspace{-1\baselineskip}
    \caption{QP state migration.}
    \label{fig-qp-migration}
    \vspace{-1.5\baselineskip}
\end{figure}
To enable retransmission at the breakpoint (i.e., the last chunk failed to be sent), it is crucial to synchronize and migrate the states from the primary QP to the backup QP.
As shown in Figure \ref{fig-qp-migration}, \X maintains three pointers to track transmission and reception states at both the sender and receiver, along with a sender-side \texttt{SyncFifo} buffer for state synchronization (Table \ref{tab:syncfifo} in Appendix \ref{appendix-structure-syncfifo}).
At the sender side, \textit{posted} tracks chunks prepared by the GPU, \textit{transmitted} marks chunks for which the CPU proxy has invoked \texttt{ibv\_post\_send}, and \textit{acked} indicates the receipt of a WC acknowledged by the receiver.
Correspondingly, the receiver tracks \textit{posted} (ready for reception), \textit{received} (invoked via \texttt{ibv\_post\_recv}), and \textit{done} (data successfully committed to the application buffer).
The receiver’s \textit{done} pointer is synchronized with the sender’s \textit{acked} pointer upon the completion of each chunk.
Additionally, the \texttt{SyncFifo} stores the metadata required for in-place retransmission: \texttt{fifoHead} manages CTS synchronization, while \texttt{restartPos} and \texttt{error \allowbreak Port} allow the sender rank to identify the resumption point and the specific faulty link, respectively.

Upon perceiving a failure, the receiver actively retreats \textit{received} to \textit{done} to make backup QP start receiving data at the breakpoint.
Then the receiver pushes requested objects to sender's buffer by backup QP to ensure the consistent breakpoint at both sides.
With mutual agreement of breakpoint location, the sender also retreats \textit{acked} and \textit{transmitted} to \texttt{restartPos}, notifying the Proxy to restart transmission from \textit{acked} by backup QP.


\noindent\textbf{Recovery of normal QPs.}
When RNIC port downs are fixed, \X migrates active transmissions from the backup QP back to the primary QP.
Because a link failure forces all associated primary QPs into the \textit{error} state, a full state-machine reset is required.
To bypass the significant overhead of QP re-establishment and bootstrap-based synchronization, we cache QP metadata (e.g., QPN) during the initial setup.
We also observe that while the control plane latency for transitioning a QP (RESET $\to$ INIT $\to$ RTR $\to$ RTS) is negligible, the underlying hardware often requires a substantial warm-up period---often on the order of seconds---to become fully operational. 
To mask this latency, \X proactively initiates the QP reset sequence immediately upon failure perception, allowing hardware preparation to overlap with the failover period. 
Consequently, once the link is restored, \X can seamlessly leverage the previously described synchronization protocol to failback to the primary QP.

\vspace{-0.5\baselineskip}
\subsection{Online Network Performance Monitor}
\label{subsec-profiling}

Although it is non-trivial to realize fine-grained network observability in current monitoring mechanism (\S\ref{subsec-challenge}), fortunately, the narrow-waist abstraction of RDMA programming \cite{kong2022collie, kim2019freeflow} can help us capture $O(\mu s)$ online inter-host network dynamics of each RNIC.
Meanwhile, by modeling fine-grained online network bandwidth, we can also locate transient link anomalies for proactive isolation before they directly affect training or cause fatal failures, as Appendix \ref{appendix-additional-experience} shows.

\begin{figure}[t]
  \centering
  \subfigure[Per-message scheme.]{\includegraphics[width=0.43\textwidth]{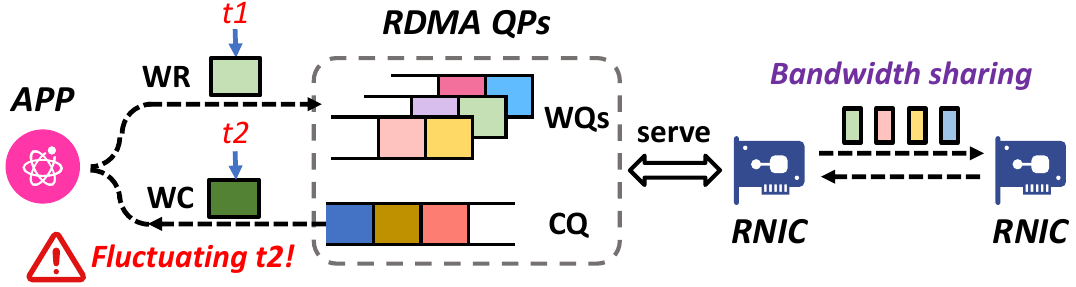}}
  \subfigure[Per-window scheme.]{\includegraphics[width=0.43\textwidth]{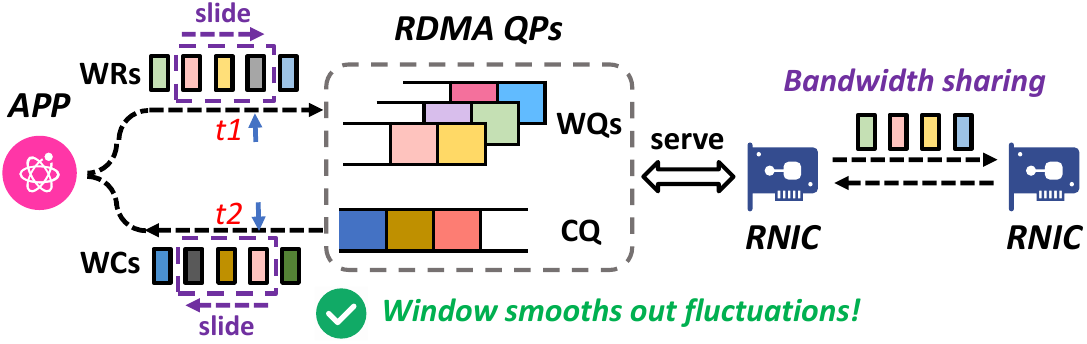}}
  \vspace{-1\baselineskip}
  \caption{Bandwidth estimation.}
  \label{fig-verb-estimation}
  \vspace{-1.5\baselineskip}
\end{figure}

\noindent\textbf{Capturing online network performance.}
To monitor fine-grained CC performance, a straightforward approach is to estimate per-message throughput from WR/WC timestamps, as shown in Figure~\ref{fig-verb-estimation}(a). 
When the application posts a WR to a WQ, \X records timestamp $t_1$; when the corresponding WC is generated on the CQ, \X records $t_2$. 
The throughput of message $M$ is then estimated as $B=\frac{\omega(M)}{t_2-t_1}$, where $\omega(M)$ is the message size.
However, per-message estimation is unstable under concurrent traffic. 
When multiple messages share the same physical link, $t_2-t_1$ captures not only transmission time, but also queuing delay and bandwidth interleaving with other requests. 
As a result, the estimated throughput becomes sensitive to transient cross-traffic rather than reflecting persistent changes in link behavior, making it unsuitable for robust anomaly detection.

\X instead uses a sliding window to estimate average throughput, as shown in Figure~\ref{fig-verb-estimation}(b). 
For each window $W$, \X records $t_1$ as the posting time of the first WR in the window and $t_2$ as the completion time of the last corresponding WC. 
It then computes the window throughput as $\overline{B} = \frac{\sum_{i \in W}{\omega(M_i)}}{t_2-t_1}$, where $i \in W$ indicates the $i^{th}$ message inside the window $W$.
By aggregating multiple messages, the window-based estimator amortizes transient queuing effects and provides a more stable signal for detecting sustained throughput degradation.

In practical usage, a proper window size is crucial to measuring accuracy.
A window that is too small fails to sufficiently dampen transient noise, resulting in the same volatility observed in the naive per-message scheme. 
Conversely, an excessively large window suppresses short-term variations but may smooths out subtle but significant throughput shifts caused by actual network anomalies and obscuring transient dynamics.
Therefore, developers should optimize the size of sliding window according to their cluster environment and network dynamics.

\noindent\textbf{Network anomaly pinpointing.}
Beyond providing $O(\mu s)$-level network monitoring, \X can effectively identify network anomalies by leveraging the window-based monitor. 
Given the inherent performance variance across different CC primitives, we employ an operation-aware heuristic. 
Specifically, \X compares the current bandwidth with the average preceding bandwidth (e.g., 10 ms).
To mitigate false positives, an anomaly is identified only when two conditions are met simultaneously: 
(i) the current bandwidth drops by more than 50\% relative to the previous average of the same primitive, and 
(ii) the remaining unsent data on the NIC (tracked via RDMA WR/WC lifecycle) exceeds twice the maximum value observed within the historical data. 
This dual-threshold strategy ensures precise localization of port-level degradation (see \S\ref{subsec-eval-monitor} for details).
Notably, thresholds are empirically chosen to avoid false positives from cases like overlapping communication, following prior works \cite{deng2025mycroft}. 

%% file: texfiles/5deployment_2.tex
\section{Deployment: \X in production}
\label{sec-deployment}

\X has been built on multiple versions of NCCL, with about additional 7,000 lines of C codes, which is publicly available at Github \cite{vccl-github}.
\X was initially deployed in a 2k-GPU cluster in December 2023, and has since been integrated into all our subsequent delivery clusters, scaling up to 24k GPUs.
To protect our customer's privacy, specific institution identities are omitted.
Originally, \X was developed for customized optimizations, like topology-aware channel reconstruction or enhanced NIC selection in dual-port environments.
Since NCCL has introduced some similar capabilities during evolution, we only focus on the deployment of \X regarding the newly proposed features.

\noindent\textbf{Cluster environment.}
To ensure the reproducibility of our experimental results, we describe an operational 24k-GPU cluster composed of multiple 3–4k-GPU sub-clusters. 
It includes: 
1) 24k NVIDIA Hopper GPUs; 
2) NVIDIA ConnectX-7 RNICs \cite{cx7} and BlueField-3 DPUs \cite{bf3datasheet}. 
Each server is equipped with 8 homogeneous GPUs and 9 RNICs. 
Each sub-cluster adopts a 1:1 oversubscribed two-layer rail-optimized network topology with 400Gbps bandwidth, and all sub-clusters are jointly managed via Kubernetes-based scheduling \cite{luksa2017kubernetes} as a unified production cluster.

\begin{figure}[t]
  \centering
  \subfigure[Inter-node P2P bandwidth.]
  {\vspace{-1.5\baselineskip} \includegraphics[width=0.48\linewidth]{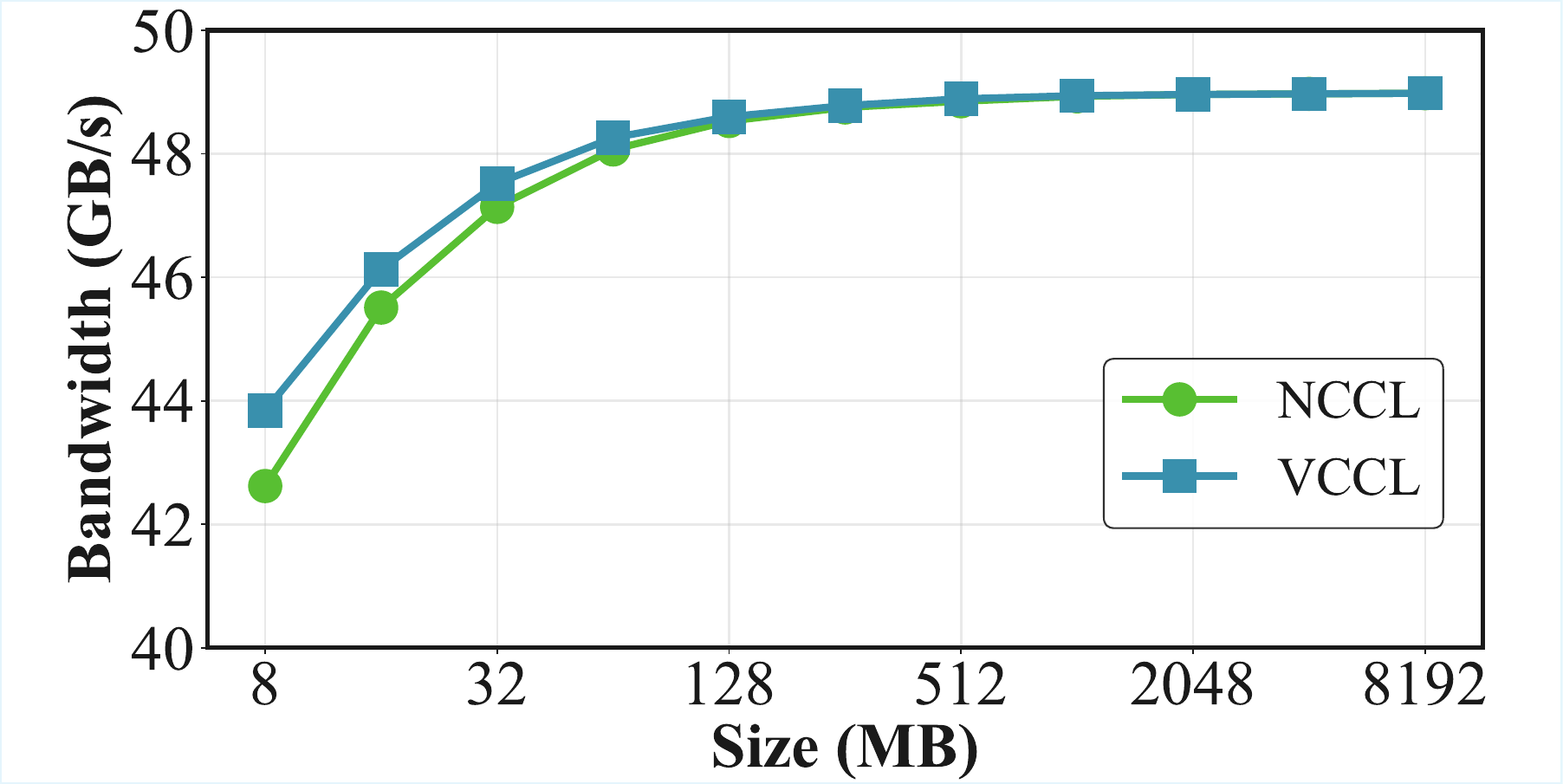}}
  \subfigure[Inter-node P2P latency.]{\vspace{-1.5\baselineskip} \includegraphics[width=0.48\linewidth]{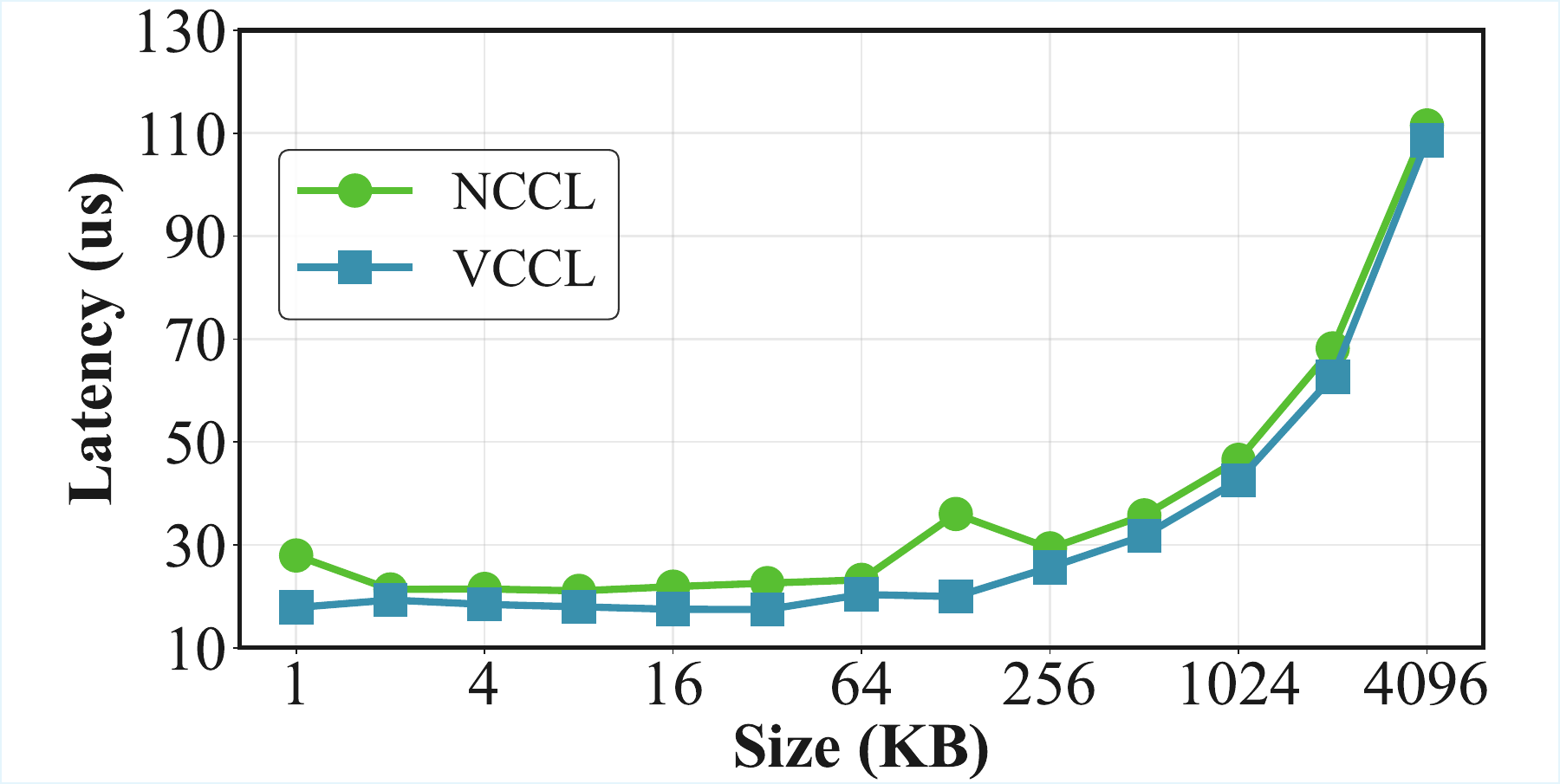}} 
  \\
  \subfigure[Intra-node P2P bandwidth.]{\vspace{-1.5\baselineskip} \includegraphics[width=0.48\linewidth]{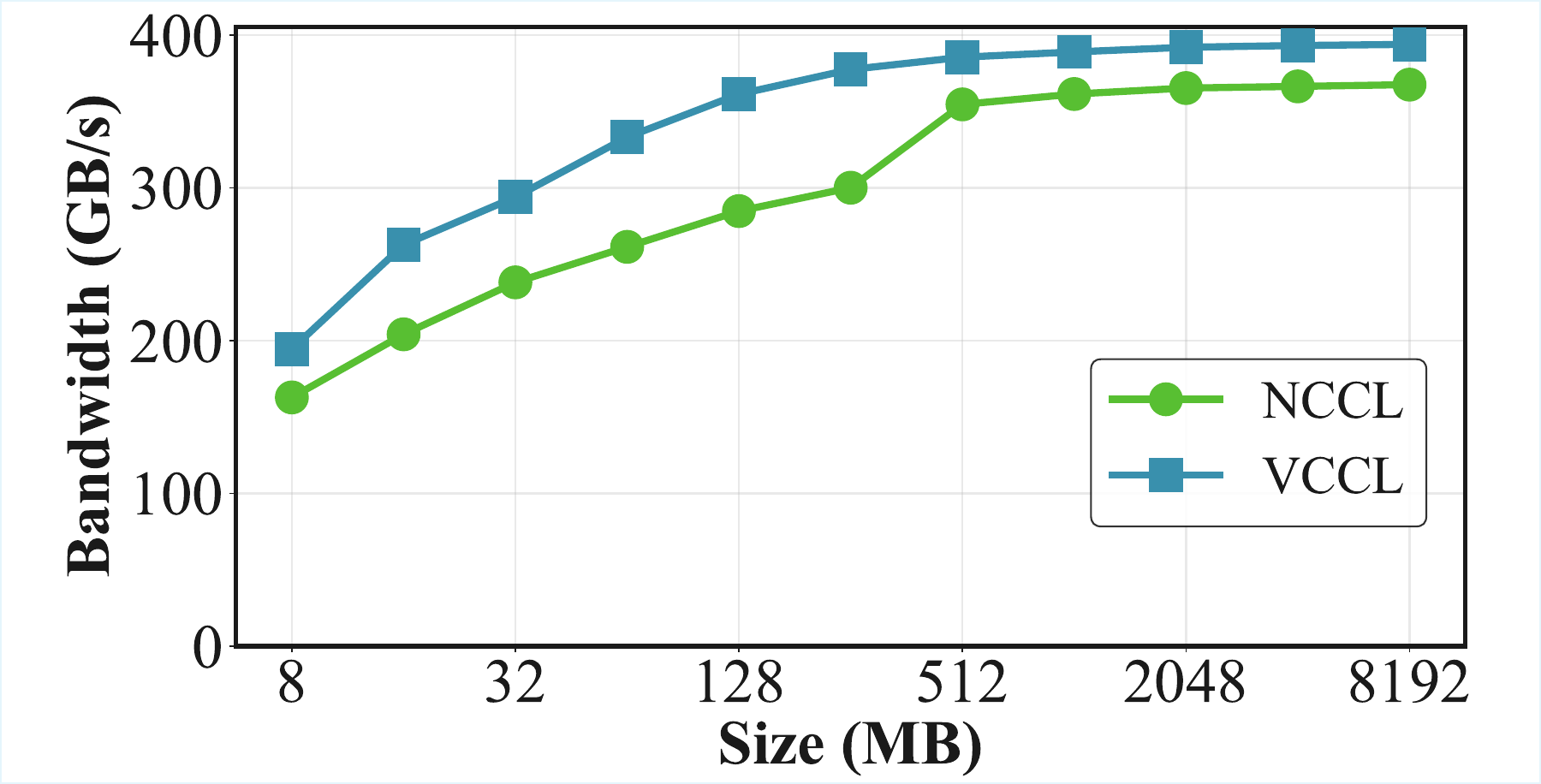}}
  \subfigure[Intra-node P2P latency.]{\vspace{-1.5\baselineskip} \includegraphics[width=0.48\linewidth]{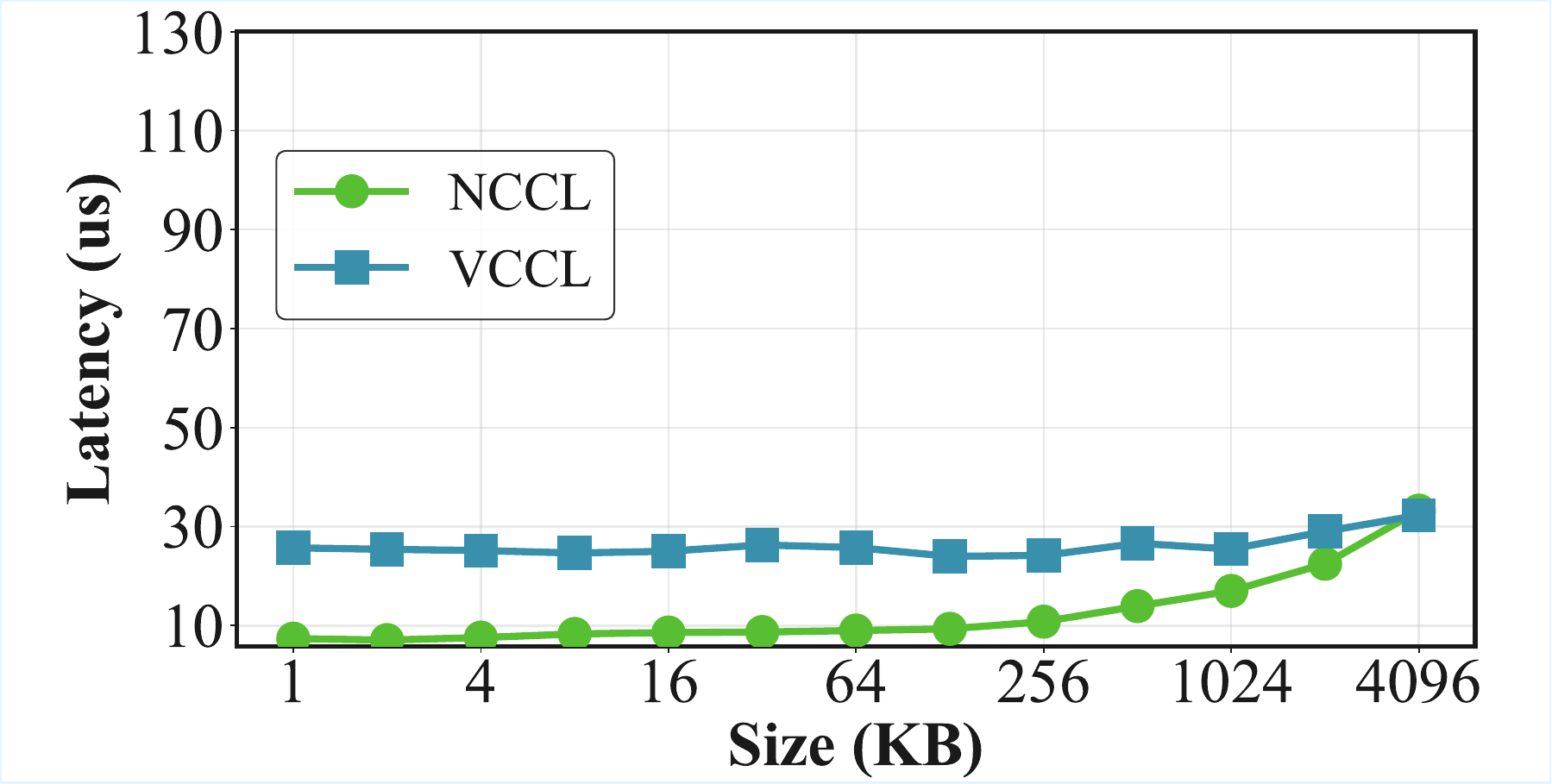}}
  \vspace{-1\baselineskip}
  \caption{Bandwidth and latency performance.}
  \label{fig-p2p-performance}
  \vspace{-1.5\baselineskip}
\end{figure}

\vspace{-0.5\baselineskip}
\subsection{Efficiency}
\label{subsec-eval-efficiency}

By achieving perfect overlap between computation and communication, \X accelerates training progress, saving great GPU hours.
Due to confidential reasons, the statistics showed below are all collected by task from our internal team.

\noindent\textbf{P2P performance.}
We first use NCCL-Tests to demonstrate the P2P performance difference between \X and NCCL v2.26 \cite{nccl}.
To ensure a fair comparison, we explicitly implement the zero-copy mechanism for the NCCL baseline.
Experimental data is depicted in Figure \ref{fig-p2p-performance}.
For inter-node P2P, since \X and NCCL are both free of cumbersome buffer copies, they achieve similar bandwidth for large messages (> 8MB).
Meanwhile, \X reduces small-message latency by 18.9\% on average by eliminating GPU-CPU synchronization.
For intra-node P2P, \X improves large-message bandwidth by approximately 7\%.
The reason behind is that its copy-engine-based approach issues wider transactions that better saturate NVLink compared to SM-based transfers.
However, we also observe higher latency for small intra-node messages---a result of contention for limited copy-engine resources---this is not a critical bottleneck in production environments. 
In PP, message sizes typically exceed 32MB (explained in Appendix \ref{appendix-pp-size}), rendering small-message latency overhead negligible. 
Furthermore, the primary value of \X’s SM-free mechanism lies in eliminating SM contention, which maximizes the potential for computation-communication overlap. 
Consequently, the bandwidth efficiency and SM availability gains far outweigh the minor latency penalties for small P2P messages.

\begin{figure}[t]
  \centering
  \subfigure[177B model.]{\includegraphics[width=0.49\linewidth]{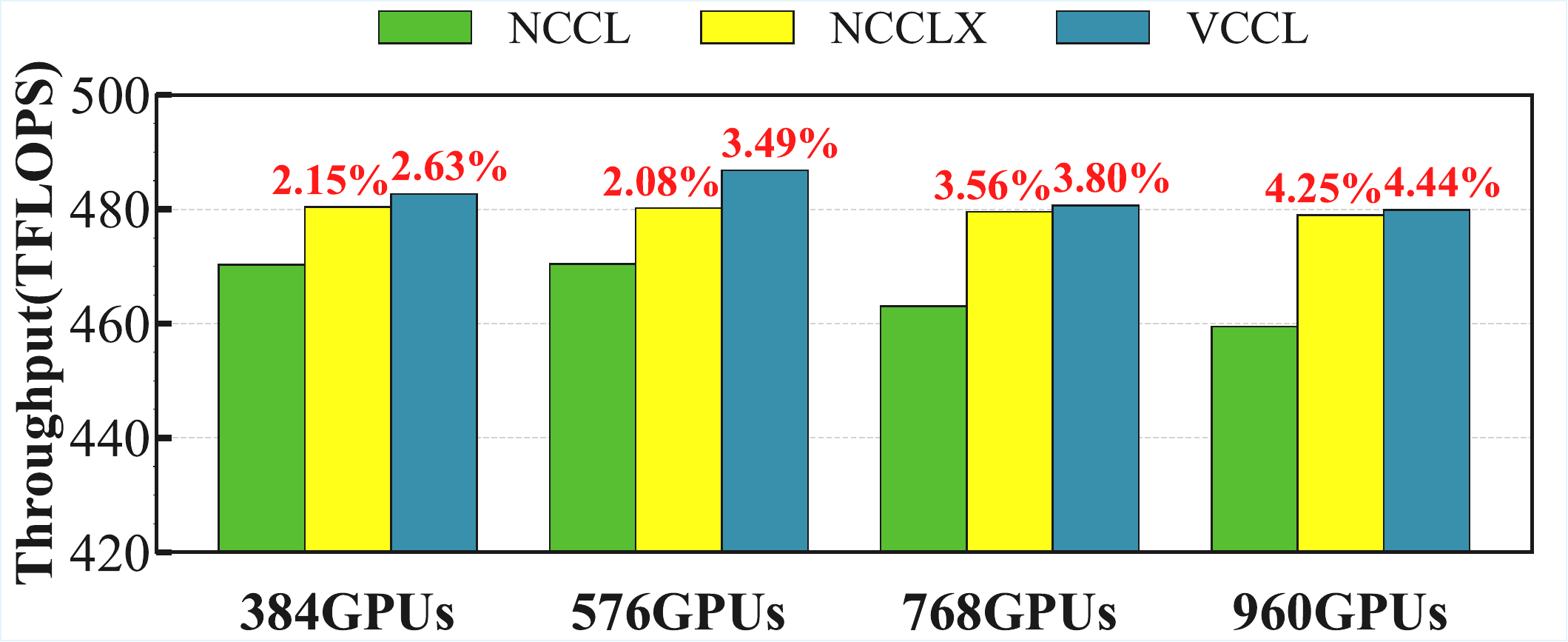}}
  \subfigure[314B model]{\includegraphics[width=0.49\linewidth]{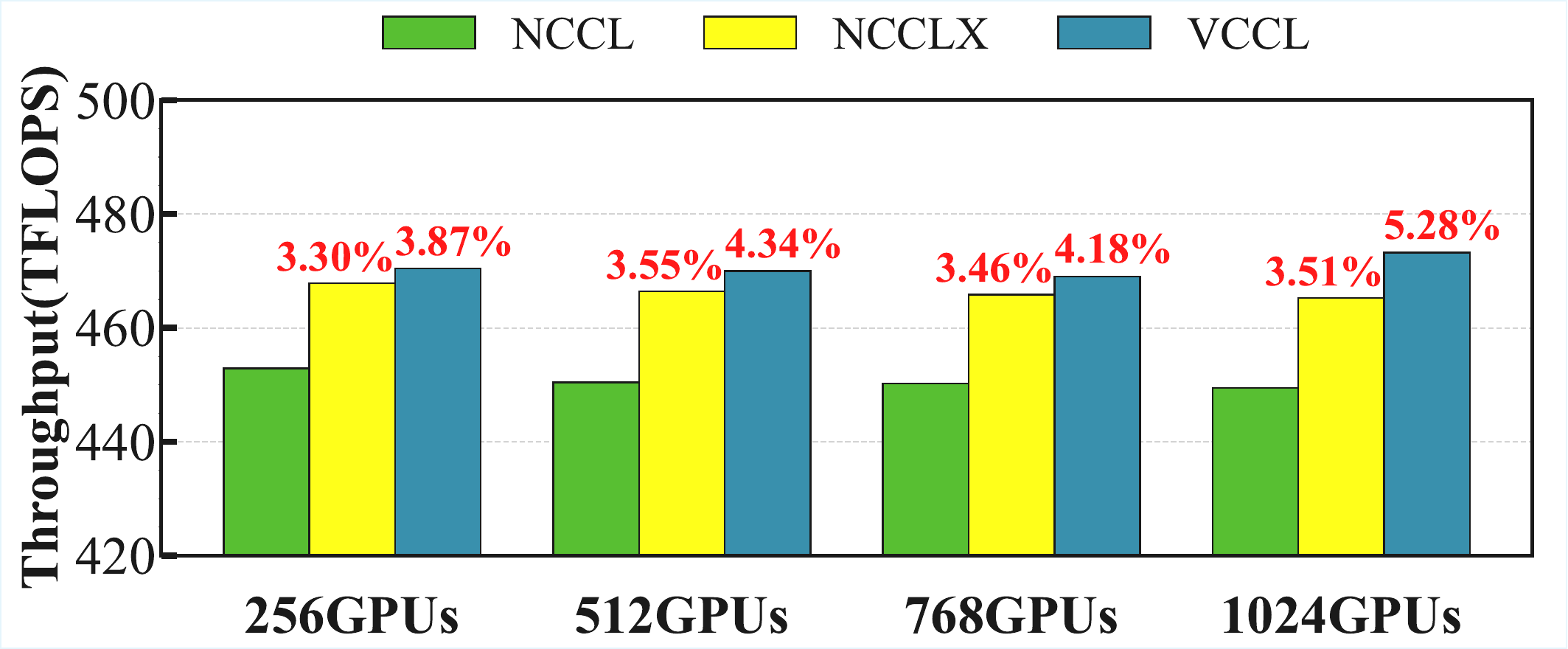}}
  \vspace{-1\baselineskip}
  \caption{Training throughput.}
  \label{fig-training-tflops}
  \vspace{-1.5\baselineskip}
\end{figure}

\noindent\textbf{Training performance.}
We then collect statistics from our internal team's training tasks to evaluate the training performance of \X.

\noindent \underline{1) Scalability. } 
We first evaluate the strong-scaling performance of \X against NCCL v2.26 and an NCCLX-like baseline that reproduces NCCLX’s key mechanism (i.e., launching a lightweight GPU kernel with only 1 SM to enforce stream ordering), but implemented in our codebase for integration and measurement consistency.
Typically, the training workload is GPT-2 \cite{radford2019language}, an open-source dense LLM with different sizes of 177B and 314B.
All our training tasks use Megatron-LM \cite{megatronlm}, with training hyperparameters and default settings listed in Appendix \ref{appendix-experiment-settings}.
As shown in Figure~\ref{fig-training-tflops}, \X consistently outperforms NCCL across all settings, delivering an average training TFLOPS improvement of 4.00\% and a maximum gain of up to 5.28\%.
Additionally, the NCCLX-based baseline exhibits performance degradations of up to 1.73\% compared to \X, indicating that even the consumption of a single SM can measurably reduce training efficiency. 
At scale, such efficiency losses translate into substantial GPU time wastage during large-scale training.
We explain the 5\% training throughput gain in Appendix \ref{appendix-sm-free-performance-gain}.

\begin{figure}[t]
  \centering
  \subfigure[32B model with 40G dataset.]{\includegraphics[width=0.49\linewidth]{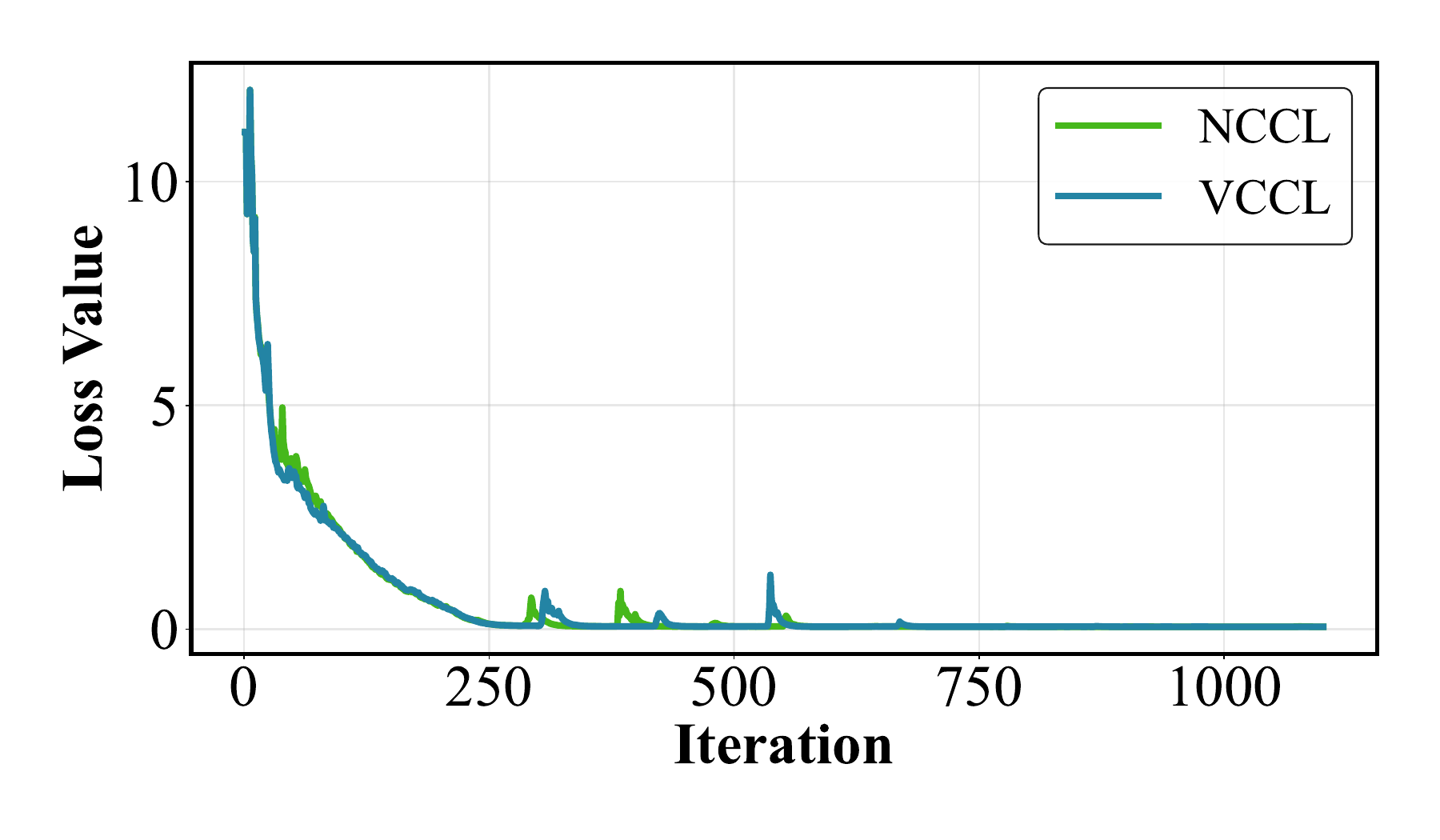}}
  \subfigure[70B model with 400M dataset.]{\includegraphics[width=0.49\linewidth]{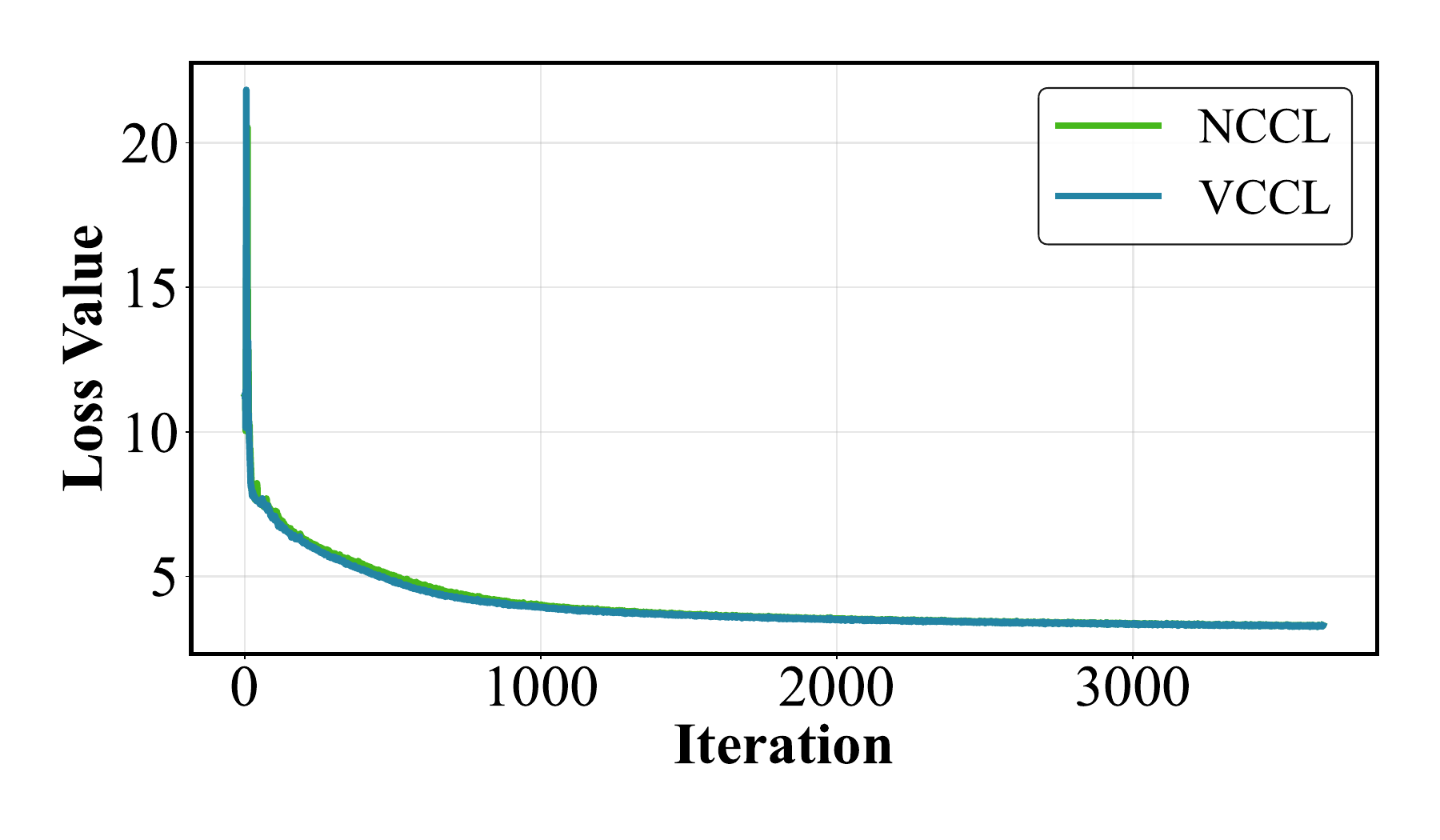}}
  \vspace{-1\baselineskip}
  \caption{Loss values.}
  \label{fig-training-loss}
  \vspace{-1.5\baselineskip}
\end{figure}

\noindent \underline{2) Model convergence. }
To verify that the SM-free design of \X does not degrade convergence speed, we conduct extensive training of GPT-2 with 32B and 70B parameters.
Figure \ref{fig-training-loss} shows that \X exhibits the same loss value trend as NCCL for GPT-2 in different sizes, proving that \X's SM-free P2P design using \texttt{waitValue} guarantees the correct interaction and order between computation and communication as kernels do in NCCL.

\noindent\textbf{Resource consumption.}
We profile the SM and CPU resources consumption of \X and NCCL, indicating that \X successfully eliminates GPU kernel launches for communication, while introducing only about 2\% additional CPU utilization (details can be seen in Appendix \ref{appendix-resource-consumption}).

\vspace{-0.5\baselineskip}
\subsection{Reliability}
\label{subsec-eval-fault-tolerance}

\begin{figure}[t]
  \centering
  \subfigure[Single-port failures.]{\includegraphics[width=0.49\linewidth]{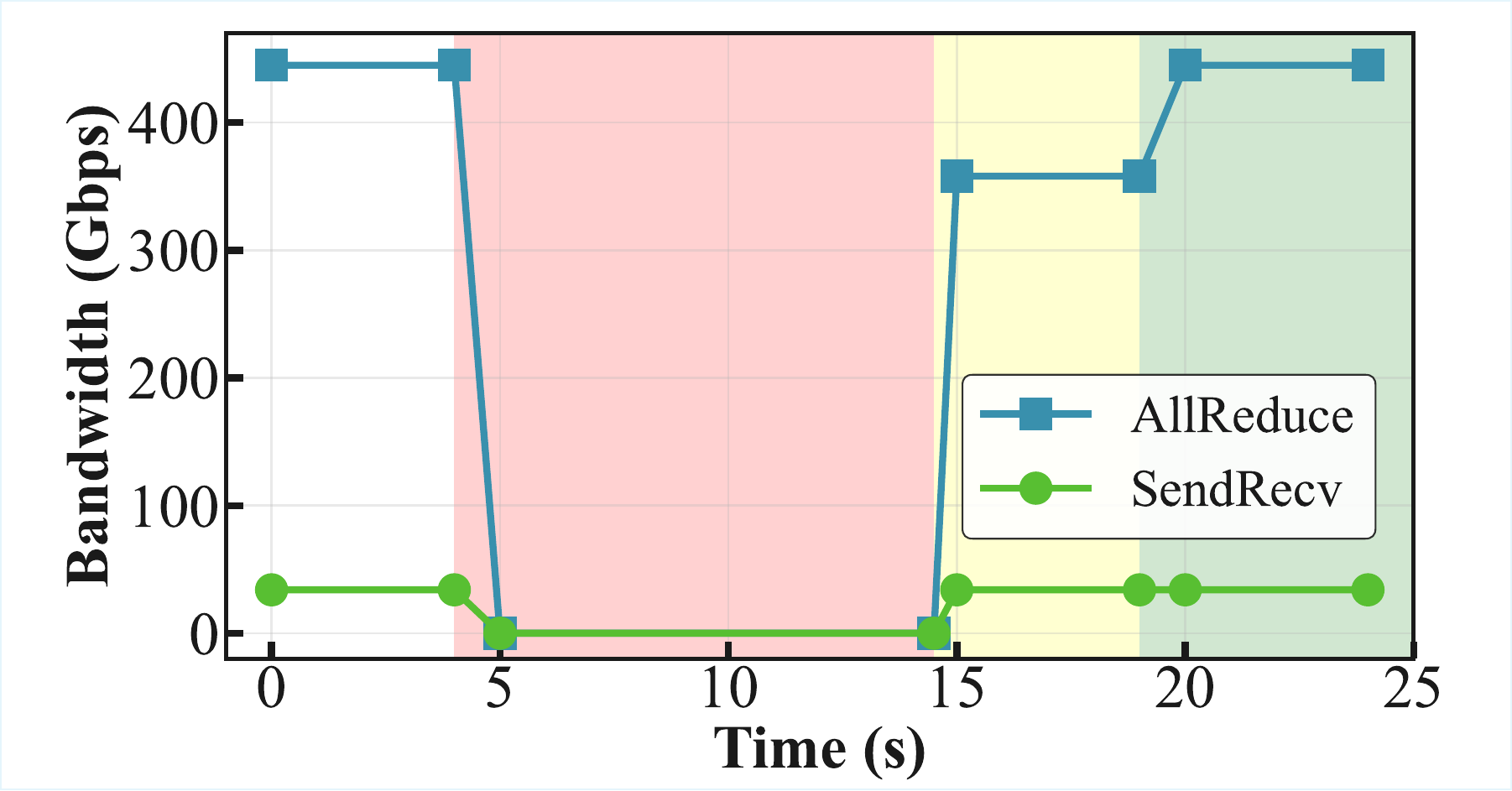}}
  \subfigure[Training performance.]{\includegraphics[width=0.49\linewidth, trim=5pt 0 5pt 0, clip]{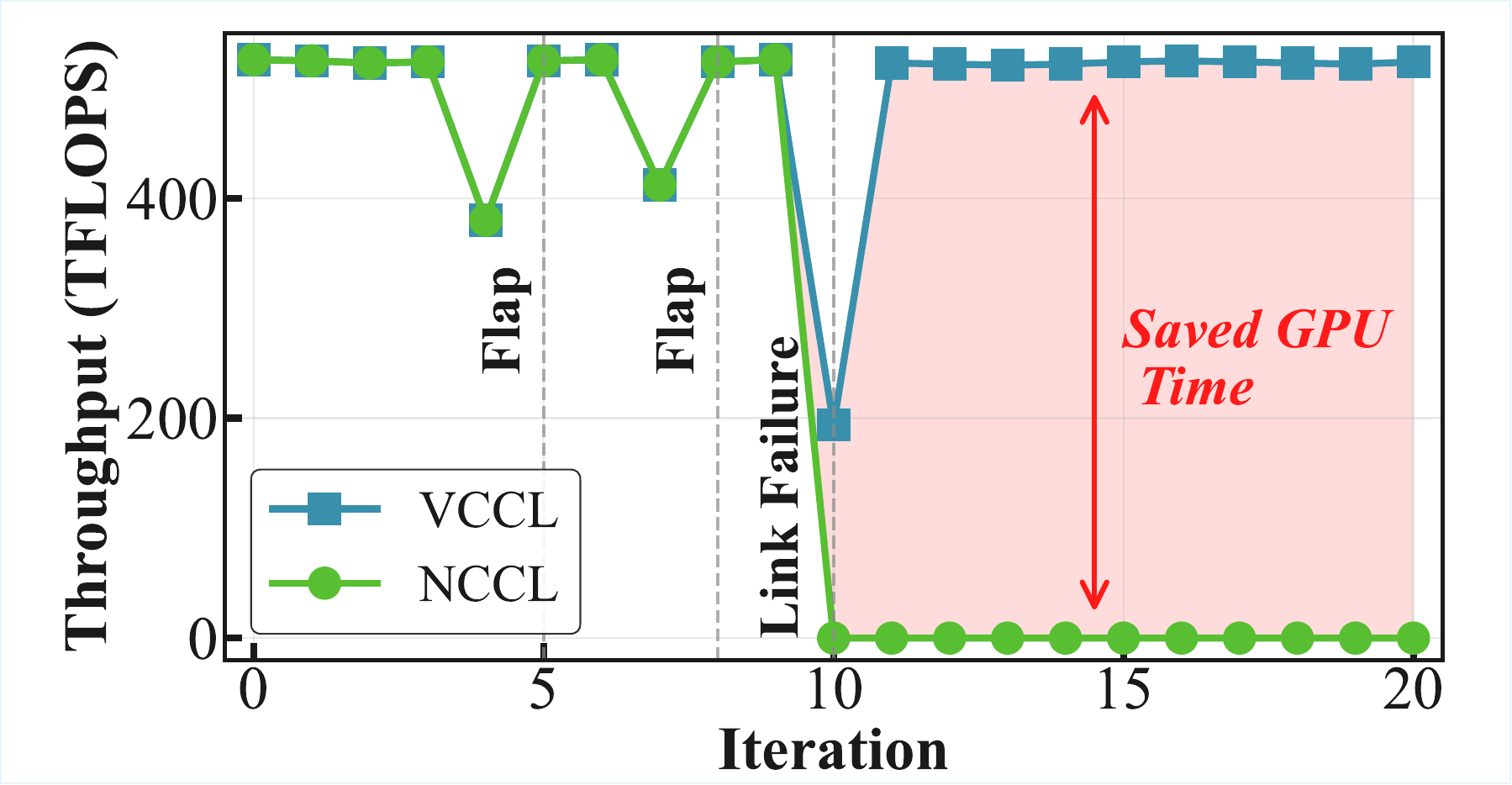}}
  \vspace{-1\baselineskip}
  \caption{Tolerance to link failures.}
  \label{fig-primary-back-qp-eval}
  \vspace{-1\baselineskip}
\end{figure}

\noindent\textbf{Tolerance to link failures.}
To evaluate the reliability of \X, we first conduct NCCL-Tests of two representative communication workloads: \texttt{Send\allowbreak Recv} (primitive for pipeline parallelism) and \texttt{All\allowbreak Reduce} (widely used in tensor and data parallelism).
Figure \ref{fig-primary-back-qp-eval}(a) depicts the runtime bandwidth of \X under a link failure.
Specifically, we manually bring down a RNIC port from 4s to 19s, whose failed duration exceeds the default retry timeout, and bring up the RNIC port at 19s again.
The red area denotes the retry period, and yellow area denotes \X is transmitting using the backup QP, while green area denotes \X switches back to the primary QP when the RNIC port is up again.
When a port down occurs, \X spends about 10s to allow hardware retransmission on the same port, staying at 0GB/s during the retry period.
After the retry window, \X switches to the backup QP using an alternative healthy RNIC port, ensuring in-place communication recovery. 
Additionally, when the RNIC port is brought up again, \X can switch back to the primary QP and deliver the ideal bandwidth.
Meanwhile, to evaluate the resilience of \X against multi-port failures, we perform an \texttt{AllReduce} stress test, where detailed results are provided in Appendix \ref{appendix-multi-port-failures}.

In addition, we select a real-world 70B-parameter training task to demonstrate \X’s robustness under practical failure conditions.
As is depicted in Figure \ref{fig-primary-back-qp-eval}(b), while both systems experience temporary stalls during minor link flaps, NCCL inevitably hangs when facing severe link failures that exceed hardware retransmission limits. 
In contrast, \X leverages its primary-backup QP mechanism to recover training after a timeout, with TFLOPS almost remaining constant. 
The red region represents the significant GPU hours \X saves. 
In large-scale production deployments, where link failures requiring manual intervention are increasingly frequent, \X’s ability to maintain continuity offers substantial savings in computational resources.


\begin{figure}[t]
    \includegraphics[width=0.9\linewidth]{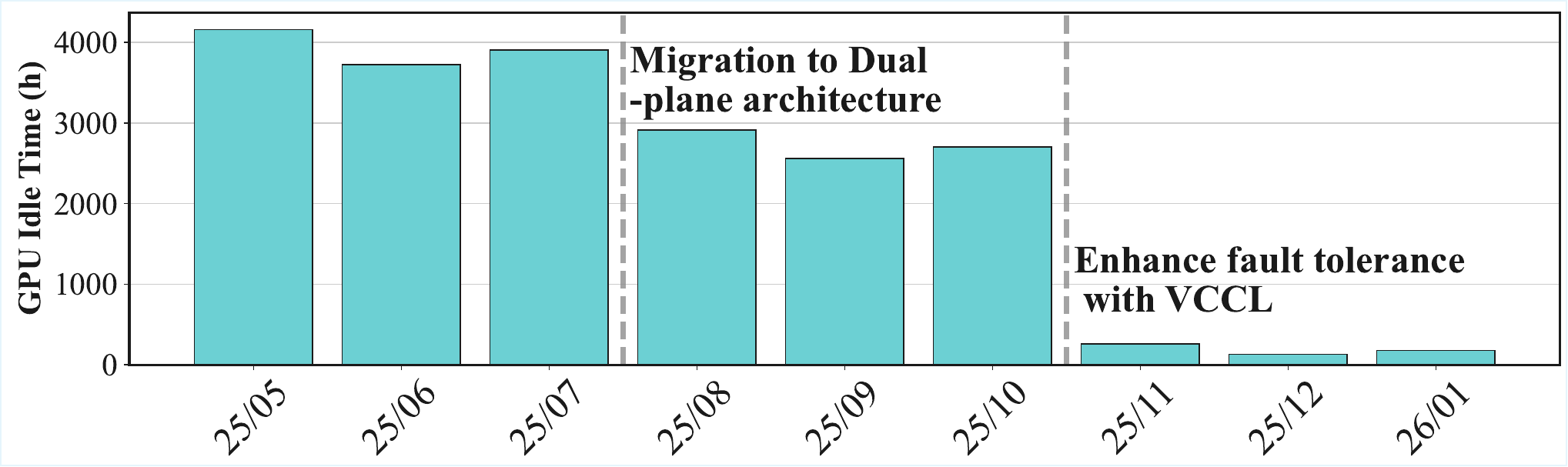}
    \centering
    \vspace{-0.5\baselineskip}
    \caption{GPU idle time caused by link failures.}
    \label{fig-production-fault-tolerance}
    \vspace{-1.5\baselineskip}
\end{figure}

\noindent\textbf{The necessity of link failure tolerance.}
We share the impact of link failures within a production cluster of 24K GPUs in Figure \ref{fig-production-fault-tolerance}.
Prior to August 2025, the cluster utilize a single-plane topology where link failures resulted in massive GPU-hour wastage. 
Although some dual-port NICs were migrated to a dual-plane deployment, reducing about 29.6\% idle GPU time, standard NIC bonding is inherently inapplicable to single-plane architecture, which still hosts the vast majority of GPUs in the cluster.
The subsequent deployment of \X's fault tolerance feature in October successfully addressed this issue, leading to a nearly 90\% reduction in idle GPU resources.
This substantial decrease in computational overhead underscores the critical necessity of runtime link-failure tolerance in large-scale distributed training clusters.

\noindent\textbf{System overhead.}
Although \X maintains twice as many QPs, active QP numbers are unchanged, so RNIC on-chip SRAM pressure does not increase. 
Meanwhile, in a 1024-GPU setting with TP=4, PP=4, and 16 channels, each RNIC needs only 192 local connections, far below modern RNIC capacity, e.g., ConnectX-7 supports 131,072 QPs.

\subsection{Observability}
\label{subsec-eval-monitor}


\begin{figure}[t]
  \centering
  \subfigure[Normal.]{\includegraphics[width=0.49\linewidth, trim=20pt 0 45pt 0, clip]{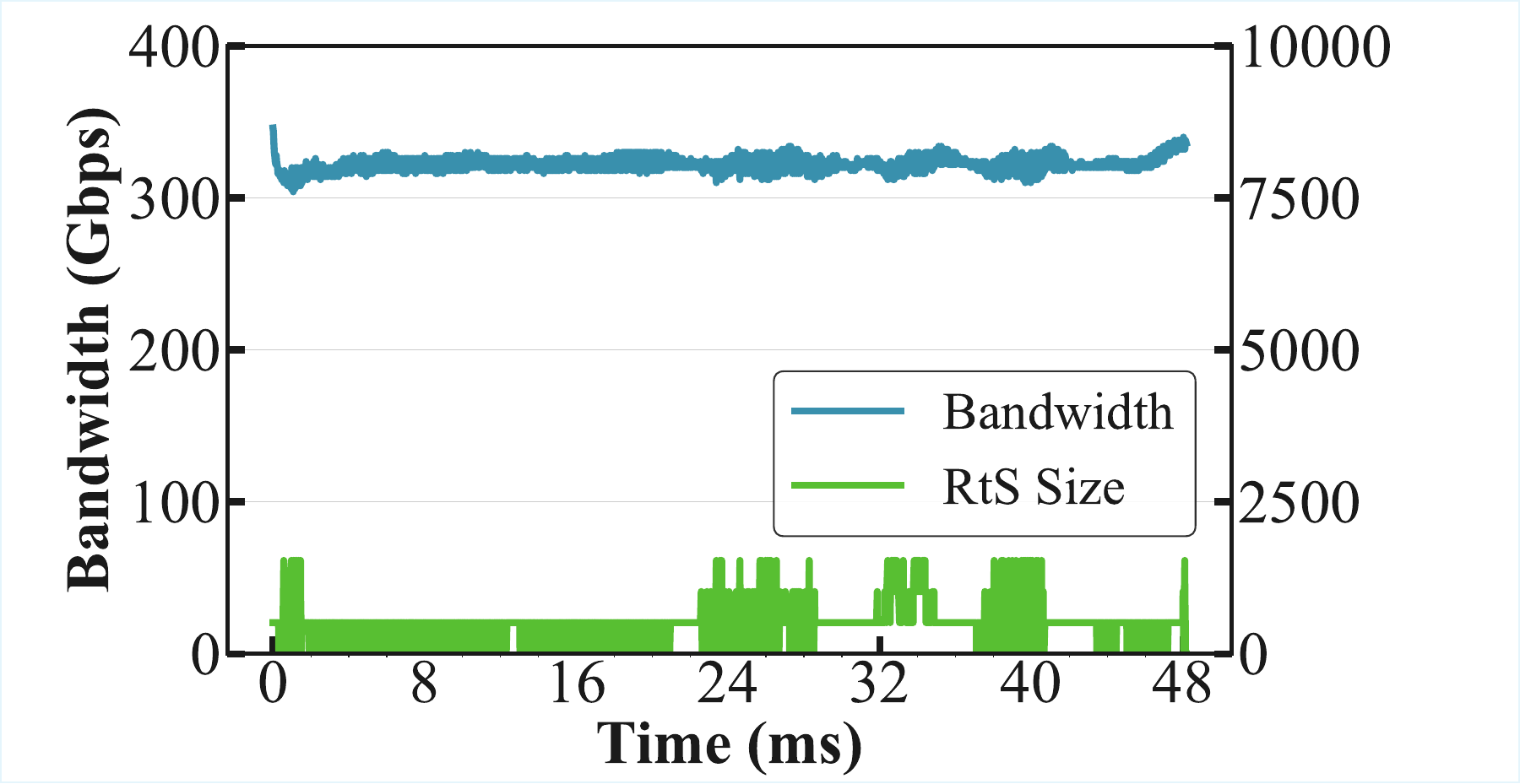}}%
  \subfigure[Manual shutdown.]{\includegraphics[width=0.49\linewidth, trim=45pt 0 20pt 0, clip]{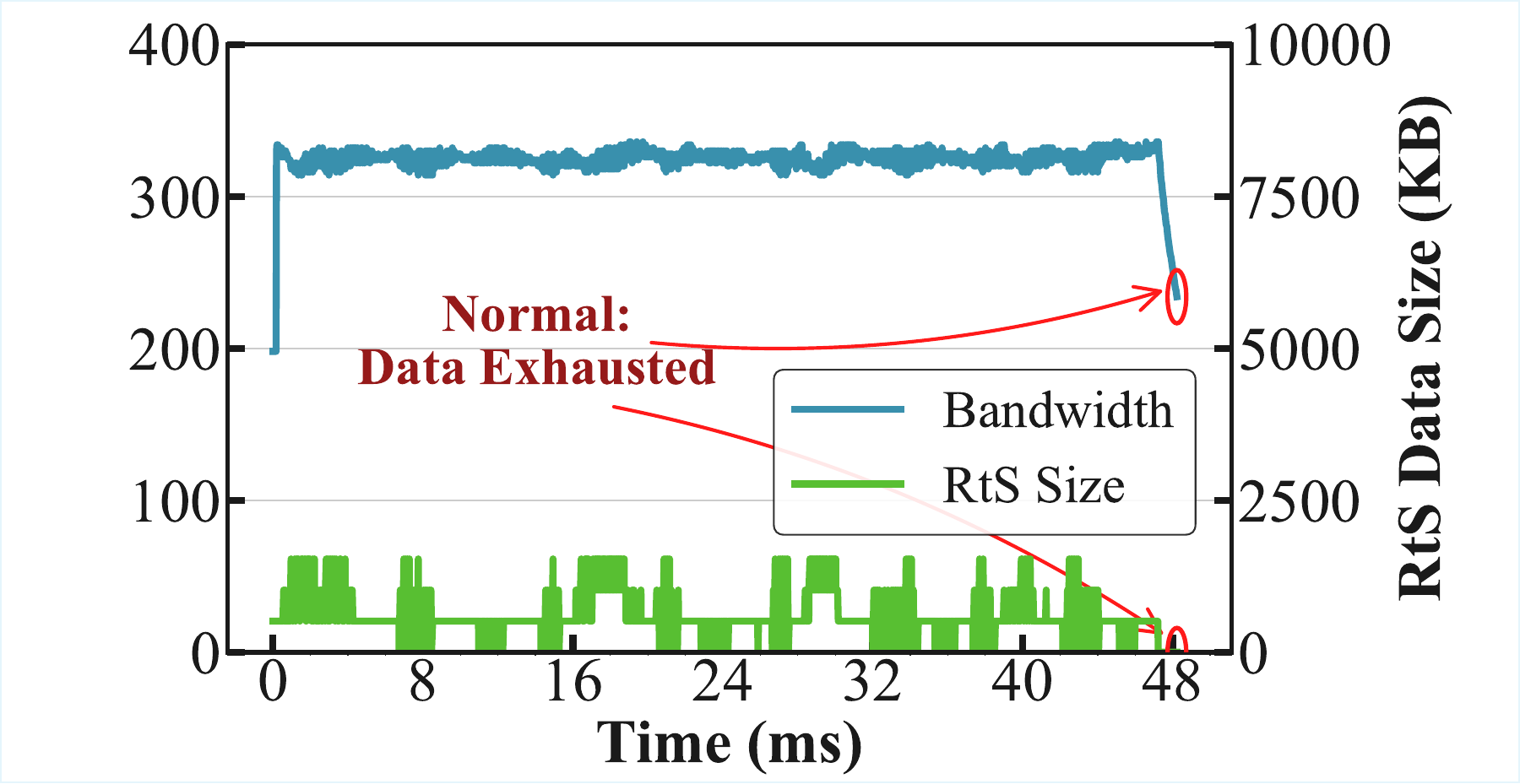}}%
  \\
  \vspace{-1\baselineskip}
  \subfigure[Background traffic.]{\includegraphics[width=0.49\linewidth, trim=20pt 0 45pt 0, clip]{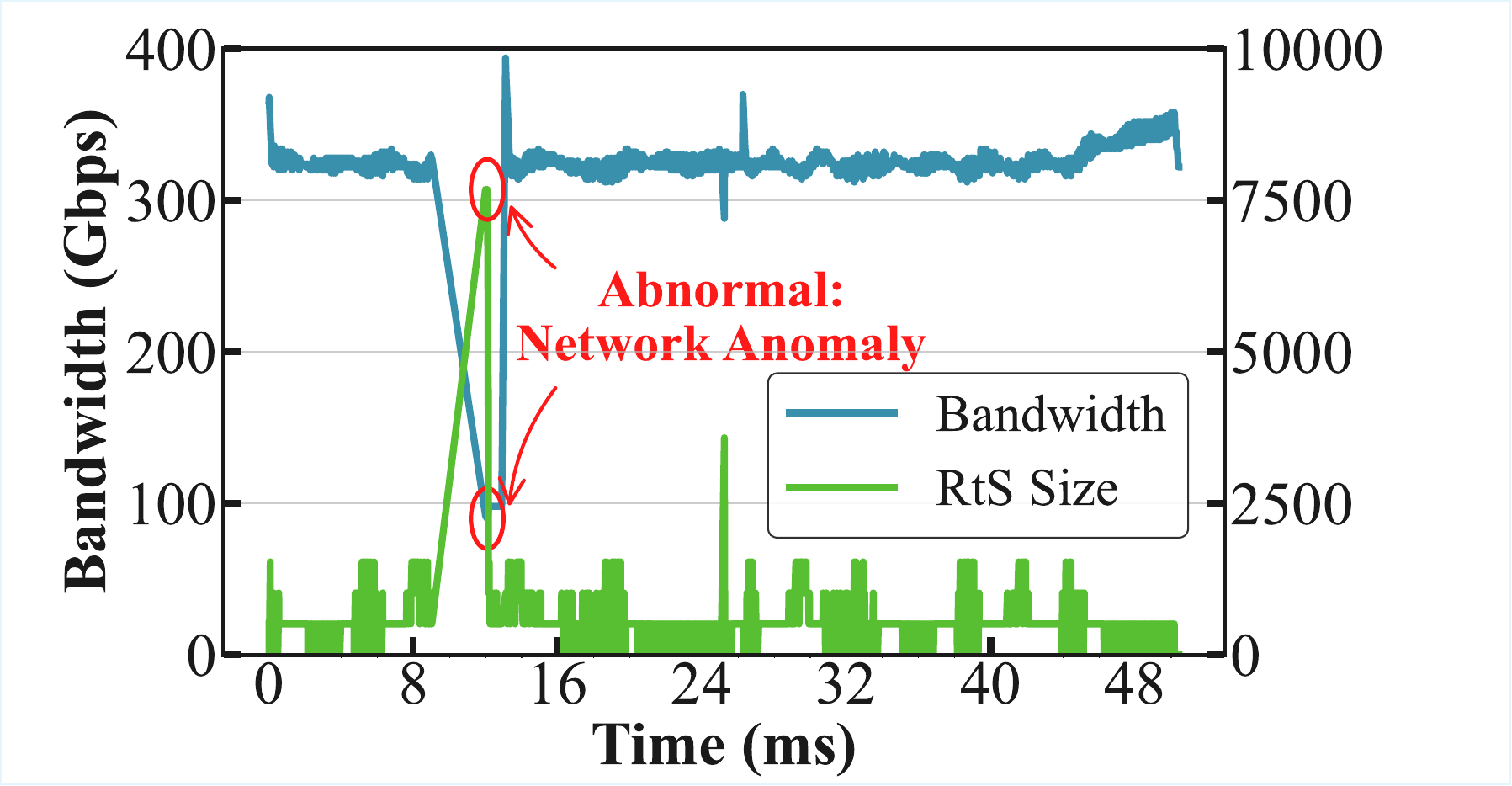}}%
  \subfigure[Background computation.]{\includegraphics[width=0.49\linewidth, trim=45pt 0 20pt 0, clip]{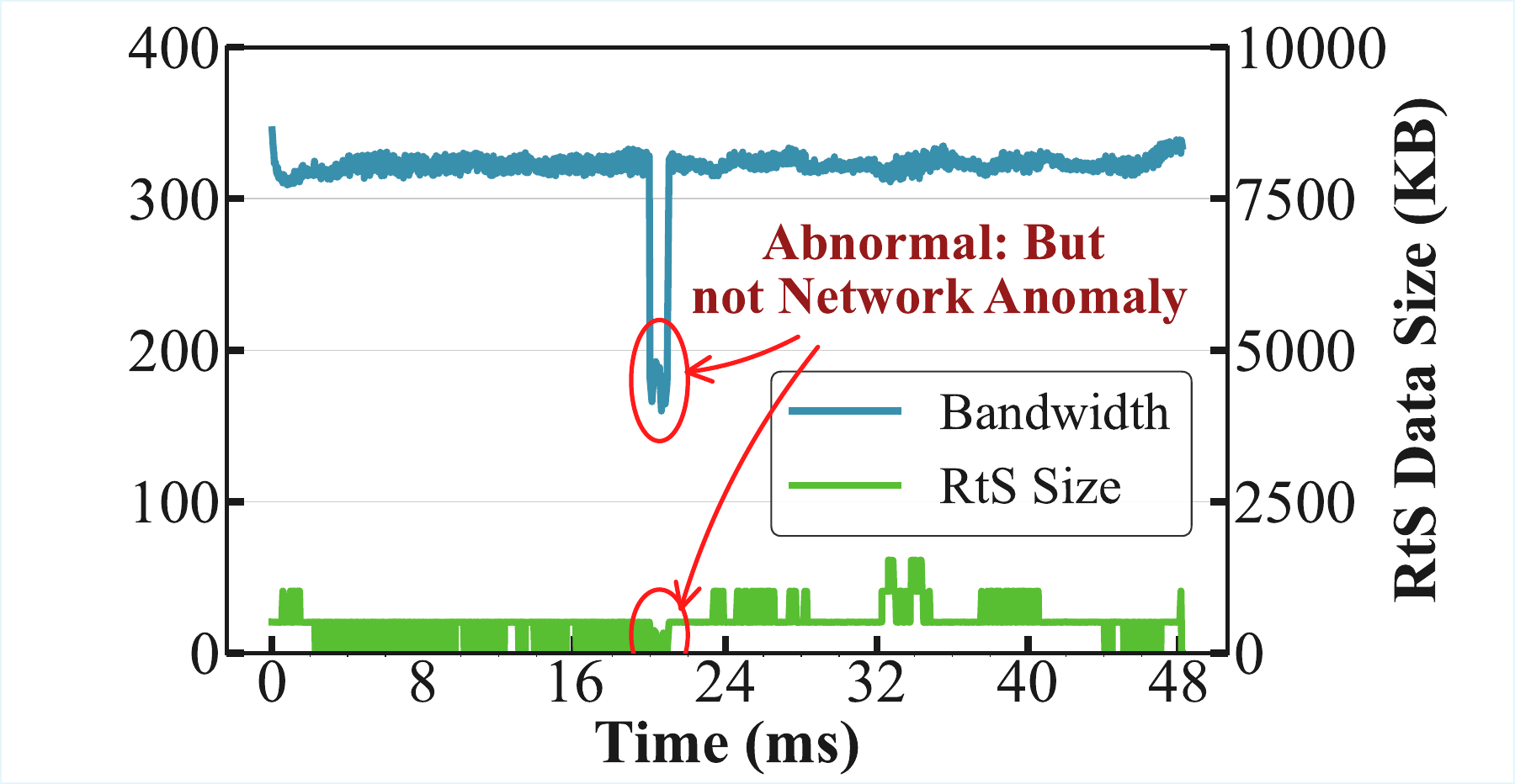}}%
  \vspace{-1\baselineskip}
  \caption{Network performance tracing of CC across four representative cases.}
  \label{fig-pinpoint}
  \vspace{-1.5\baselineskip}
\end{figure}

\noindent\textbf{Case study: Pinpointing network straggler.}
Building on the $O(\mu s)$ monitor (due to space limit, we detail the wisdom of sliding-window based monitor in Appendix \ref{appendix-sliding-window}), \X enables precise localization of network stragglers.
Figure \ref{fig-pinpoint} evaluates the effectiveness of \X’s heuristic pinpointing algorithm under four representative scenarios: 
(1) a normal CC task, 
(2) the manual termination of a CC task, 
(3) a CC task under network interference (small-packet perftest), and 
(4) a CC task under transient hardware interference (GPU-burn). 
As shown in the results, the normal case (case 1) maintains stable \texttt{bandwidth} and \texttt{remaining-to-send} (\texttt{RTS}), indicating a healthy state. 
In case 2, \X correctly classifies the terminal bandwidth decline as normal behavior, attributing it to the exhaustion of the NIC’s data buffer rather than a network anomaly.
Crucially, \X distinguishes between network and non-network issues.
In case 3, the simultaneous drop in \texttt{bandwidth} and \texttt{RTS} accumulation allows \X to pinpoint a network anomaly.
By contrast, in case 4, despite the significant \texttt{bandwidth} drop caused by GPU resource contention, the lack of data accumulation on the NIC prevents a false positive. 
These results underscore \X’s ability to accurately localize network stragglers while remaining robust to other performance fluctuations.

\begin{figure}[t]
    \includegraphics[width=0.9\linewidth]{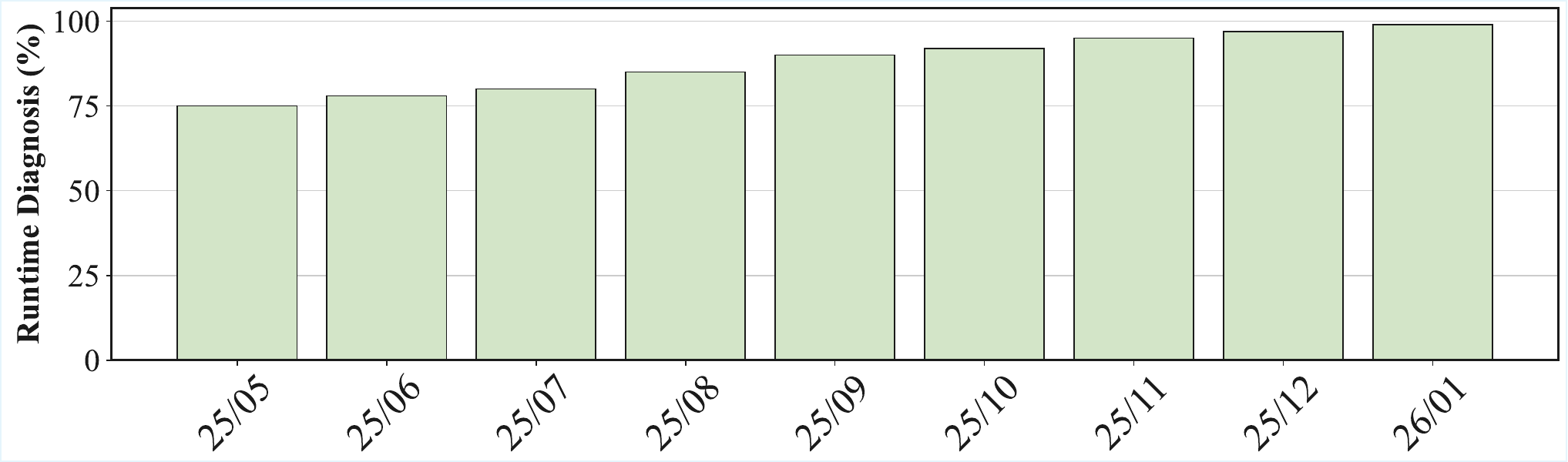}
    \centering
    \vspace{-1\baselineskip}
    \caption{Runtime diagnosis percentage.}
    \label{fig-production-troubleshooting}
    \vspace{-1.5\baselineskip}
\end{figure}

\noindent\textbf{Full-stack troubleshooting with \X.}
To enable effective runtime diagnosis, we propose a full-stack troubleshooting system with \X designed for fault localization and recovery, the details of which can be seen in Appendix \ref{appendix-full-stack-troubleshooting}.
We share the runtime diagnosis percentage in one of our operating cluster.
As depicted in Figure \ref{fig-production-troubleshooting}, with the integration of \X for network straggler pinpointing, our full-stack troubleshooting platform completes its final missing component and drives the runtime diagnosis percentage gradually approaching 100\%.
Given the high GPU wastage of offline diagnosis, these real-world results demonstrate the effectiveness of our full-stack troubleshooting platform.

\noindent\textbf{System overhead.}
We evaluate the system overhead of our window-based monitor in terms of CPU and memory.
As shown in Table \ref{tab-monitor-overhead} in Appendix \ref{appendix-resource-consumption}, it incurs only moderate CPU overhead and negligible memory footprint.

\subsection{Other System-level optimizations}

\noindent\textbf{Optimizing memory usage.}
During the deployment of \X, we observe unexpected high GPU memory consumption, reaching 10GB HBM for specific MoE models.
This inefficiency, which severely constrains model parallelism strategies, stems from NCCL's default aggressive pre-allocation across all protocols and channels, as well as its reliance on intermediate GPU buffers for intra-node data movement.
To mitigate this, \X introduces a dynamic memory pool design.
First, \X adopts a lazy allocation strategy, where resources are assigned to connections only upon their first runtime usage. 
The system maintains a 2MB-aligned memory pool to allocate minimal necessary memory for each peer, dynamically expanding via \texttt{cuCallocAsync} when the pool is exhausted. 
Second, by leveraging zero-copy mechanisms for both intra-node and inter-node P2P primitives, we are allowed to eliminate intermediate buffers, thereby minimizing the memory footprint.
Our experimental data shows that \X achieves up to 26.7\% memory reduction in practical deployments of models with complex parallelism.
More details can be found in Appendix \ref{appendix-memory-optimization}.

\noindent\textbf{Traffic imbalance in bonding environment.}
When enabling \X in dual-plane clusters with bonding, we encounter severe traffic load imbalance.
Specifically, in dual-port environments without bonding enabled, a two-node \texttt{ReduceScatter} benchmark in NCCL-Tests achieves up to 380 Gbps bandwidth, whereas enabling bonding reduces the bandwidth to approximately 220 Gbps.
To mitigate this, after deep investigation of Mellanox driver internals, we utilize \texttt{mlx5dv\_\allowbreak modify\_\allowbreak qp\_\allowbreak lag\_port} during the initialization phase to evenly distribute QPs across the two physical ports.
This approach successfully enables the bonding environment to achieve bandwidth performance similar to that of a single-port one.

%% file: texfiles/6operating.tex
\section{Experience and lessons}
\label{sec-operating}

In this section, we share key experiences and lessons we learned during the deployment of \X (additional experience can be found in Appendix \ref{appendix-additional-experience}).


\noindent\textbf{Performance gain of SM-free.}
While \S \ref{subsec-eval-efficiency} demonstrates the significant performance gain of \X's SM-free design, we encounter several unexpected behaviors in deployment.
First, occasionally, we observe that \X exhibits little or no advantage over NCCL.
Further investigation reveals that this phenomenon stems from a small batch size that leads to underutilization of GPU, allowing NCCL to also overlap communication in PP and thus masking the benefit of saving SM resources.
After we increase the batch size to fully saturate the GPU, the issue is resolved.
Besides, we find that relative throughput gain of \X gradually decreases as the training cluster scales out.
After careful investigation, we point out this is not caused by scalability limitations of \X, but by growing DP communication overhead.
The performance benefit can be modeled as: $I = \frac{\frac{C}{T_v + \alpha} - \frac{C}{T_n + \alpha}}{\frac{C}{T_n + \alpha}} = \frac{T_n - T_v}{T_v + \alpha}$, where $C$ represents TFLOPS per GPU per iteration, $T_v$ and $T_n$ denote per-iteration compute time under \X and NCCL, respectively, and $\alpha$ represents DP communication overhead.
Specifically, since the communication pattern within DP group follows the \texttt{ring} algorithm over a single-rail interconnect, communication overhead of \texttt{AllReduce} or \texttt{ReduceScatter} \& \texttt{AllGather}  exhibits linear scaling, causing $I$ to decrease with cluster size.
Despite this, even though the relative TFLOPS gain decreases at larger scale, the increased GPU count preserves significant absolute GPU time savings from \X's SM-free design.



\noindent\textbf{Misleading cases without troubleshooting.}
Without a full-stack troubleshooting platform, heterogeneous root cau\allowbreak ses frequently collapse into indistinguishable CC-level exceptions, making root-cause pinpointing ambiguous.
We present two typical cases that lead to CC throughput degradation.
\textit{1) Case 1: insufficient CPU cores.}
In this case, CC throughput only achieved 30\% of ideal bandwidth.
Subsequent investigation reveals that this issue stems from the misconfiguration of setting an upper limit of at most 8 CPU cores per developer, in which case \X's network proxy did not have enough CPU resources to conduct communication.
However, by using \textit{lscpu} on the platform, we can indeed easily pinpoint the root cause.
\textit{2) Case 2: misconfiguration of GPU fan speed.}
We encountered intermittent drops during LLM training, and Nsight \cite{nsight} analysis showed that other metrics excluding communication worked well as usual.
Further investigation identifies a fan speed limit misconfiguration on a single GPU, such that it was overheated and slowed down the communication, resulting in compromised CC throughput.
In fact, this misleading case can also be effectively diagnosed through hardware-level inspection with \textit{ipmitool}.

\noindent\textbf{Lessons we learned as a CCL provider.}
In some specific cases, our customers encountered failures that the troubleshooting platform reported no hardware or software anomalies.
However, they observed these failures could be eliminated simply by switching from \X to NCCL, which naturally leads to the perception that the issues were caused by \X. 
Through careful investigation, we find that these failures are often not inherent bugs of \X, but rather stem from subtle deployment issues.
Below, we present two representative lessons we learned in real-world deployments.

\noindent \textit{1) Environment variable misconfigurations can cause fatal failures.} 
After deploying \X, some customers experienced severe segmentation faults. 
Notably, these failures disappeared when reverting to NCCL, making the issue particularly misleading. 
We initially analyzed the core dumps using \texttt{gdb}, yet no obvious software defects or memory violations attributable to \X were identified.
A deeper inspection of the training scripts eventually revealed the root cause: the environment variable \texttt{ICCL\_NET\_PLUGIN} was mistakenly set, forcing \X to load NCCL’s default network plugin. 
This resulted in conflicting definitions of internal data structures between \X and NCCL, ultimately causing undefined behavior and crashes. 
Once this environment variable was unset, \X operated correctly.
This case highlights that, as a CCL provider, careful recommendation of environment variables is essential. 
Even seemingly small configuration errors can introduce subtle and catastrophic failures that are difficult to diagnose through conventional debugging tools.

\noindent \textit{2) Rule out NCCL bugs before attributing failures to \X.}
We encountered a more subtle failure scenario in which \X consistently crashed while NCCL functioned correctly. 
This difference initially suggested a defect in \X.
Further investigation revealed that the issue was related to MTU inconsistency.
At the time, \X was based on NCCL v2.21.5, which did not yet include software-side mitigations for handling MTU mismatches.
In contrast, the NCCL plugin bundled in the customer’s runtime environment was NCCL v2.24, where this issue had already been fixed. 
As a result, NCCL appeared to function correctly, while \X, built on the older version, inherited the bug and failed.
This experience suggests that, when diagnosing suspected failures of \X, it is critical to first align the NCCL versions used by both \X and the runtime environment. 
Excluding NCCL-induced bugs early in the debugging process helps avoid misattribution of faults and allows more effective improvement of \X’s robustness and performance.

\noindent\textbf{Robust networking configuration.}
Networking configuration is critical to \X's reliability, yet we encounter a subtle but fatal failure---unexpected changes of Global Identifier (GID). 
In RoCEv2 environments, while GIDs should remain static after initialization, we discover that default OS-level network managers (e.g., \texttt{NetworkManager}) trigger a GID increment upon RNIC port flaps even if the original IP is reclaimed. 
This non-deterministic GID shift inevitably collapses the communication collective and the LLM training job. 
While driver-level patches can suppress GID increments, they introduce compatibility risks in virtualized networks. 
Consequently,  we explore an alternative by changing the configuration method for IP addresses in a lightweight and reliable manner.
We take Linux Ubuntu operating system as an example.
Ubuntu has two basic IP configuration methods, i.e., \texttt{NetworkManager} and \texttt{Networking} service.
By \texttt{NetworkManager}, GIDs can change because of a temporary loss of IP addresses when RNIC port are down, illustrated in the aforementioned case.
Instead, when we use \texttt{Networking} to configure the IP addresses, even if port down happens, its IP address still remains unchanged and avoids a temporary loss, leading to an unchanged GID number.
Other Linux distributions can also use the similar approaches for IP address configuration.

%% file: texfiles/7discussion.tex
\section{Discussion}
\label{sec-discussion}

\noindent\textbf{SM-free and device-initiated.}
Recent industry advancements \cite{deepep, pplx-kernels} leverage device-initiated APIs (e.g., GPUDirect Async \cite{gda}) to minimize network latency by offloading control plane operations from CPU proxy to GPU kernels. 
While this approach is fundamentally distinct from \X's SM-free architecture, we argue that the two paradigms are complementary. 
Device-initiated APIs accelerate RNIC doorbell triggering, offering significant latency optimization for small messages, which is a critical factor for latency-sensitive scenarios like distributed LLM inference. 
Conversely, in training scenarios where efficiency hinges on maximizing overlapping between computation and communication rather than pure latency, \X's SM-free design is superior.

\noindent\textbf{SM-free for other reduction-free primitives.}
In this paper, we mainly focus on optimizations of P2P primitives.
The reasons behind this choice is that \X's SM-free P2P implementation enables perfect overlapping of computation and communication within the 1F1B pipeline, thereby optimizing the critical path of LLM training. 
We believe the SM-free philosophy of \X can also be extended to other reduction-free CC primitives. 
Specifically, for \texttt{AlltoAll} in Mixture-of-Experts (MoE) workloads, our SM-free design can also facilitate overlapping computation with the dispatch and combine phases. 
However, optimizing \texttt{AlltoAll} entails intricate traffic scheduling considerations to avoid many-to-one incast.
On the other hand, \texttt{AllGather} often lies on the critical path of training and lacks sufficient computation-communication overlap opportunities.
Therefore, we leave the extension to other primitives as our future work.

\noindent\textbf{Limitation of fault tolerance.}
While \X provides robust tolerance against multi-port link failures, it remains vulnerable to the simultaneous failure of both the primary and backup RNICs due to our fixed QP placement strategy. 
Fortunately, our production-level observations indicate that the probability of such correlated failures is near zero, justifying our current design choice. 
Nonetheless, we believe addressing this corner case through more flexible backup mechanisms offers a compelling direction for academic research. 
Furthermore, given the significant bandwidth fluctuations observed during multi-port failures, developing adaptive communication scheduling based on runtime hardware resource availability represents another promising area for future exploration.

%% file: texfiles/8relatedwork.tex
\section{Related Work}
\label{sec-related}

\noindent\textbf{CC Optimization.}
Besides NCCL and various xCCLs from industrial community, many works from academia are proposed to optimize CC performance.
Blink \cite{wang2020blink}, SCCL \cite{cai2021synthesizing}, TACCL \cite{shah2023taccl} and SyCCL \cite{cao2025syccl} synthesize collective algorithms according to heterogeneous topology of intra-host and inter-host.
TCCL \cite{kim2024tccl}, TECCL \cite{liu2024rethinking}, MCCS \cite{wu2024mccs}, AutoCCL \cite{xu2025autoccl}, ResCCL \cite{liu2025resccl} and Wang et al. \cite{zhao2025efficient} optimize the CC scheduling and configuration by topology, traffic workload and user information.
Despite promising, these works are not mature and comprehensive enough to address our challenges in production-level clusters.

\noindent\textbf{Production-level LLM training.}
ByteDance \cite{jiang2024megascale}, Meta \cite{gangidi2024rdma, matam2024quickupdate, grattafiori2024llama, si2025ncclx}, Alibaba \cite{qian2024alibaba, dong2025evolution}, DeepSeek \cite{an2024fire, guo2025deepseek}, Kuaishou \cite{yuan2024accelerating} and Shanghai AI lab \cite{hu2024characterization} share their practice on production-level LLM training, including optimizations on efficiency of computation \& communication, reliability and observability for their large-scale GPU training clusters.
However, they do not provide a systematic prospective on requirements, designs and deployment of collective communications.
We expect \X together with existing works can contribute to a more thriving LLM training ecosystem.

%% file: texfiles/9conclusion.tex
\section{Conclusion}
\label{sec-conclusion}

In this paper, we design \X, an efficient, reliable and observable CCL with SM-free P2P, primary-backup QPs and window-based network monitor for production-level GPU clusters.
Our production deployment and evaluation show that \X improves training performance by saving SM resources, maintains training under RNIC port failures, and detects $O(\mu s)$ network anomalies.
We also share operational lessons from running \X at scale, and hope they motivate the community to revisit CC and build better CCL service for their clusters.

%% file: texfiles/10appendix.tex
\newpage
\appendix
\section*{Appendix}

\section{NCCL SM utilization}
\label{appendix-nccl-sm-utilization}

Table \ref{tab-nccl-sm} presents the SM utilization of different P2P workloads in NCCL.
Experimental data in Table \ref{tab-nccl-sm} shows that intra-host p2p (32SMs by default) occupies up to 13.2\% SM while inter-host p2p (2SMs by default) consumes 1.8\% SM.
Meanwhile, single-node alltoall (28SMs by default) uses up to 13.1\% SM whereas two-nodes alltoall (4SMs by default) utilizes 2.3\% of total SM.
\begin{table}[h]
    \vspace{-0.5\baselineskip}
    \caption{NCCL SM utilization of P2P workloads.}
    \centering
   \renewcommand{\arraystretch}{1.1}
    \footnotesize
    \vspace{-1\baselineskip}
    \setlength{\tabcolsep}{1.5mm}{
    \begin{tabular}{|c|c|c|c|c|}
    \hline
     \diagbox[width=9em]{Metric}{Workload} & \begin{tabular}[c]{@{}c@{}}Intra-host\\ P2P\end{tabular} & \begin{tabular}[c]{@{}c@{}}Inter-host\\ P2P\end{tabular} & \begin{tabular}[c]{@{}c@{}}8 ranks\\ \texttt{alltoall}\end{tabular} & \begin{tabular}[c]{@{}c@{}}16 ranks\\ \texttt{alltoall}\end{tabular} \\ \hline
     Default SMs & 32 & 2 & 28 & 4 \\ \hline
    SM utilization (\%) & 13.2 & 1.8 & 13.1 & 2.3 \\ \hline
    \end{tabular}
    }
    \label{tab-nccl-sm}
    \vspace{-1\baselineskip}
\end{table}




\section{Structure of \texttt{SyncFifo}}
\label{appendix-structure-syncfifo}

\texttt{SyncFifo} maintains the state required for in-place chunk retransmission. 
Table \ref{tab:syncfifo} details the structure.
Specifically, \texttt{fifoHead} supports \texttt{CTS} synchronization, while the sender rank relies on \texttt{restartPos} and \texttt{errorPort} to determine the resumption point and identify the faulty link.

\begin{table}[h]
    \centering
    \caption{Structure of \texttt{SyncFifo.}}
    \label{tab:syncfifo}
    \vspace{-1\baselineskip}
    \footnotesize
    \begin{tabular}{ll}
        \toprule
        \textbf{Object} & \textbf{Meaning} \\
        \midrule
        \texttt{fifoHead} & Synchronize the offsets of CTS message. \\
        \midrule
        \texttt{restartPos} & Represent the retransmission chunk (\textit{done}). \\
        \midrule
        \texttt{errorPort} & Indicate the abnormal port. \\
        \bottomrule
    \end{tabular}
    \vspace{-1.5\baselineskip}
\end{table}

\section{Analysis of data size in PP}
\label{appendix-pp-size}

In pipeline parallelism (PP), activations are communicated between adjacent pipeline stages. 
The size of these communicated tensors is determined by the hidden dimension, sequence length, and micro-batch size, which together often lead to communication volumes exceeding tens of megabytes.
Formally, consider a transformer-based model with hidden size $H$, sequence length $L$, and micro-batch size $B$. 
The activation tensor exchanged between two consecutive pipeline stages can be expressed as
\begin{equation}
\mathcal{A} \in \mathbb{R}^{B \times L \times H}.
\end{equation}
Assuming a data type with precision $p$ bytes (e.g., $p=2$ for FP16), the communication size per pipeline transfer is
\begin{equation}
S_{\mathrm{PP}} = B \times L \times H \times p.
\end{equation}
For modern large-scale language models, $H$ typically ranges from several thousand to over ten thousand, while $L$ is often set to hundreds or thousands. 
Even with a modest micro-batch size, the resulting communication volume quickly exceeds $32$ MB. 
For example, with $B=4$, $L=2048$, $H=8192$, and FP16 precision, we obtain
\begin{equation}
S_{\mathrm{PP}} = 4 \times 2048 \times 8192 \times 2 \approx 128 \ \mathrm{MB}.
\end{equation}
In addition, pipeline parallelism typically employs multiple micro-batches to improve device utilization. 
Let $M$ denote the number of micro-batches in a pipeline schedule. 
The cumulative communication volume per iteration then scales linearly with $M$:
\begin{equation}
S_{\mathrm{PP}}^{\mathrm{iter}} = M \times S_{\mathrm{PP}}.
\end{equation}
These factors jointly explain why inter-stage communication in PP systems frequently involves data transfers significantly larger than $32$ MB, making communication efficiency a critical bottleneck in large-scale pipeline-parallel training.

\section{Experiment Settings}
\label{appendix-experiment-settings}

The training hyperparameters and \X settings of our evaluation (\S\ref{sec-deployment}) are listed in Table \ref{tab-hyperparameter}.

\begin{table}[h]
    \caption{Training hyperparameters and \X settings.}
    \centering
    \small
    \renewcommand{\arraystretch}{1.1}
    \vspace{-0.5\baselineskip}
    \begin{tabular}{|c|c|c|}
    \hline
    Category & Setting & Value \\ \hline
    \multirow{11}{*}{Training} & Global batch size & 512 \\ \cline{2-3} 
     & Micro batch size & 1 \\ \cline{2-3} 
     & DP size & 8 \\ \cline{2-3} 
     & TP size & 2 \\ \cline{2-3} 
     & PP size & 4 \\ \cline{2-3} 
     & Attention head number & 64 \\ \cline{2-3} 
     & Sequence length & 2048 \\ \cline{2-3} 
     & layers\_per\_virtual\_pipeline\_stage & 2 \\ \cline{2-3} 
     & Learning rate & 1.5e-4 \\ \cline{2-3} 
     & Optimizer & Adam \\ \cline{2-3} 
     & Precision & BF16 \\ \hline
    \multirow{3}{*}{\X} & IB\_TIMEOUT & 18 \\ \cline{2-3} 
     & IB\_RETRY\_CNT & 7 \\ \cline{2-3} 
     & Window size & 8 \\ \hline
    \multirow{3}{*}{\begin{tabular}[c]{@{}c@{}}CC traffic\\ generator\end{tabular}} & Message size & 4G \\ \cline{2-3} 
     & QP number & 2 \\ \cline{2-3} 
     & Channel number & 32 \\ \hline
    \end{tabular}
    \label{tab-hyperparameter}
\end{table}

\section{Performance gain of SM-free}
\label{appendix-sm-free-performance-gain}
The 5\% gain stems from reduced SM contention and the elimination of scheduling-induced stragglers. 
For example, in 1F1B, the GEMM kernel \texttt{nvjet\_tst\_256x128\_64x4\_1x2\_h \allowbreak \_bz\_coopA\_TNT} launches 132 thread blocks, each with 12 warps, on a Hopper GPU with 132 SMs. 
In isolation, this kernel naturally forms one wave of execution, with roughly one GEMM block assigned to each SM. 
However, when it co-runs with the NCCL SendRecv kernel, which launches two blocks with 20 warps each, the NCCL blocks can be co-resident with GEMM blocks on two of the SMs. 
On these SMs, the 12 GEMM warps no longer have exclusive access to the SM's warp schedulers and execution resources; instead, they share and compete for scheduler issue slots, memory pipelines, and other SM resources with the 20 communication warps. 
Since the GEMM kernel cannot complete until all of its blocks finish, these few delayed GEMM blocks become tail stragglers and extend the kernel's critical path. 
\X avoids this interference by preventing the communication kernel from being scheduled onto GEMM-occupied SMs, thereby removing the straggler-induced stalls that would otherwise delay GEMM completion.



\section{Resource Consumption of \X}
\label{appendix-resource-consumption}

Table \ref{tab-sm} shows the kernel invocation and SM utilization of inter-host P2P communication using NCCL and \X.
The second column in the table represents the invoked kernels and their respective percentage of executing duration, which are collected by Nsight.
Obviously, NCCL invokes a \texttt{send/recv} kernel \texttt{ncclDevKernel\_SendRecv} to perform the reduction-free P2P communication, which accounts for 68.3\% execution time and 3.2\% SM usage.
Instead, except for two inevitable NCCL-Tests kernels (i.e., \texttt{prepareInput2} for data preparation and \texttt{verifyPrepare} for result verification), \X does not launch any GPU kernels to perform the P2P communication, thereby consuming lower SM utilization than NCCL.
Note that in practical LLM training, \texttt{prepareInput2} and \texttt{verifyPrepare} will not be invoked, such that \X's SM-free P2P can achieve almost zero SM consumption, which can be scheduled in computation to accelerate the LLM training process.

\begin{table}[h]
    \caption{Kernel invocation and SM consumption.}
    \centering
    \footnotesize
    \renewcommand{\arraystretch}{1.1}
    \vspace{-1\baselineskip}
    \begin{tabular}{|c|l|c|}
    \hline
     & \multicolumn{1}{c|}{Operations (percentage)} & SM utilization \\ \hline
    \multirow{5}{*}{NCCL} & \texttt{ncclDevKernel\_SendRecv} (68.8\%) & \multirow{5}{*}{3.2\%} \\ \cline{2-2}
    & \texttt{verifyPrepare} (27.0\%) &  \\ \cline{2-2}
     & \texttt{prepareInput2} (3.5\%) &  \\ \cline{2-2}
     & {[}CUDA memset{]} (0.4\%) &  \\ \cline{2-2}
     & {[}CUDA memcpy Host-to-Device{]} (0.4\%) &  \\ \hline
    \multirow{4}{*}{\X} & \texttt{verifyPrepare} (86.1\%) & \multirow{4}{*}{2.8\%} \\ \cline{2-2}
     & \texttt{prepareInput2} (11.4\%) &  \\ \cline{2-2}
     & {[}CUDA memset{]} (1.3\%) &  \\ \cline{2-2}
     & {[}CUDA memcpy Host-to-Device{]} (1.2\%) &  \\ \hline
    \end{tabular}
    \label{tab-sm}
    \vspace{-1\baselineskip}
\end{table}

\begin{figure}[h]
    \includegraphics[width=0.95\linewidth]{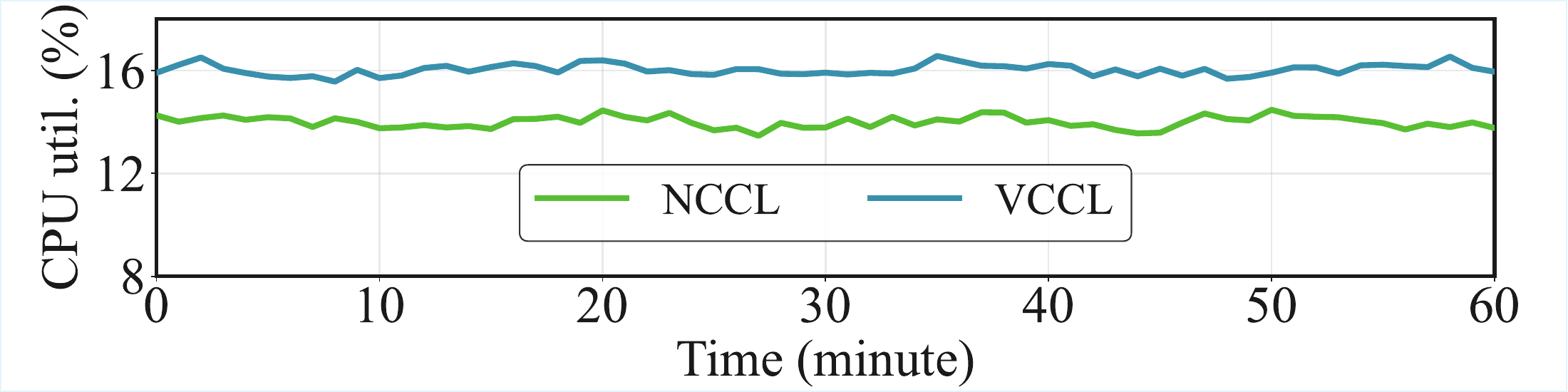}
    \centering
    \vspace{-1\baselineskip}
    \caption{CPU utilization.}
    \label{fig-kernelfree-cpu}
    \vspace{-1\baselineskip}
\end{figure}

Figure \ref{fig-kernelfree-cpu} compares the CPU utilization of a GPU server using NCCL and \X, respectively.
Although \X's SM-free design offloads P2P from GPU kernels to CPU proxies, it only introduces about 2\% more CPU utilization than NCCL.
The result validates \X's deployability and feasibility in real-world GPU cluster, where a low CPU utilization is essential to maintain the CC efficiency (discussed in \S\ref{sec-discussion}).

Table \ref{tab-monitor-overhead} reports the system overhead introduced by the online monitor. 
Enabling the monitor increases CPU utilization from 9.32\% to 21.1\%, corresponding to an additional 11.8\% CPU overhead, while memory utilization rises marginally from 1.7\% to 2.1\%. 
Overall, the online monitor incurs moderate CPU overhead and negligible memory overhead, indicating that its resource consumption remains acceptable for practical deployment.

\begin{table}[h]
    \caption{System overhead of online monitor.}
    \centering
    \footnotesize
    \renewcommand{\arraystretch}{1.1}
    \vspace{-1\baselineskip}
    \begin{tabular}{|c|c|c|}
    \hline
    \diagbox[width=11.8em]{Hardware}{Scheme} & w/o monitor & w/ monitor \\ \hline
    CPU utilization & 9.32\% & 21.1\% \\ \hline
    Memory utilization & 1.7\% & 2.1\% \\ \hline
    \end{tabular}
    \label{tab-monitor-overhead}
    \vspace{-1\baselineskip}
\end{table}

\section{Resilience under multi-port failures.}
\label{appendix-multi-port-failures}
\begin{figure}[h]
    \includegraphics[width=0.95\linewidth]{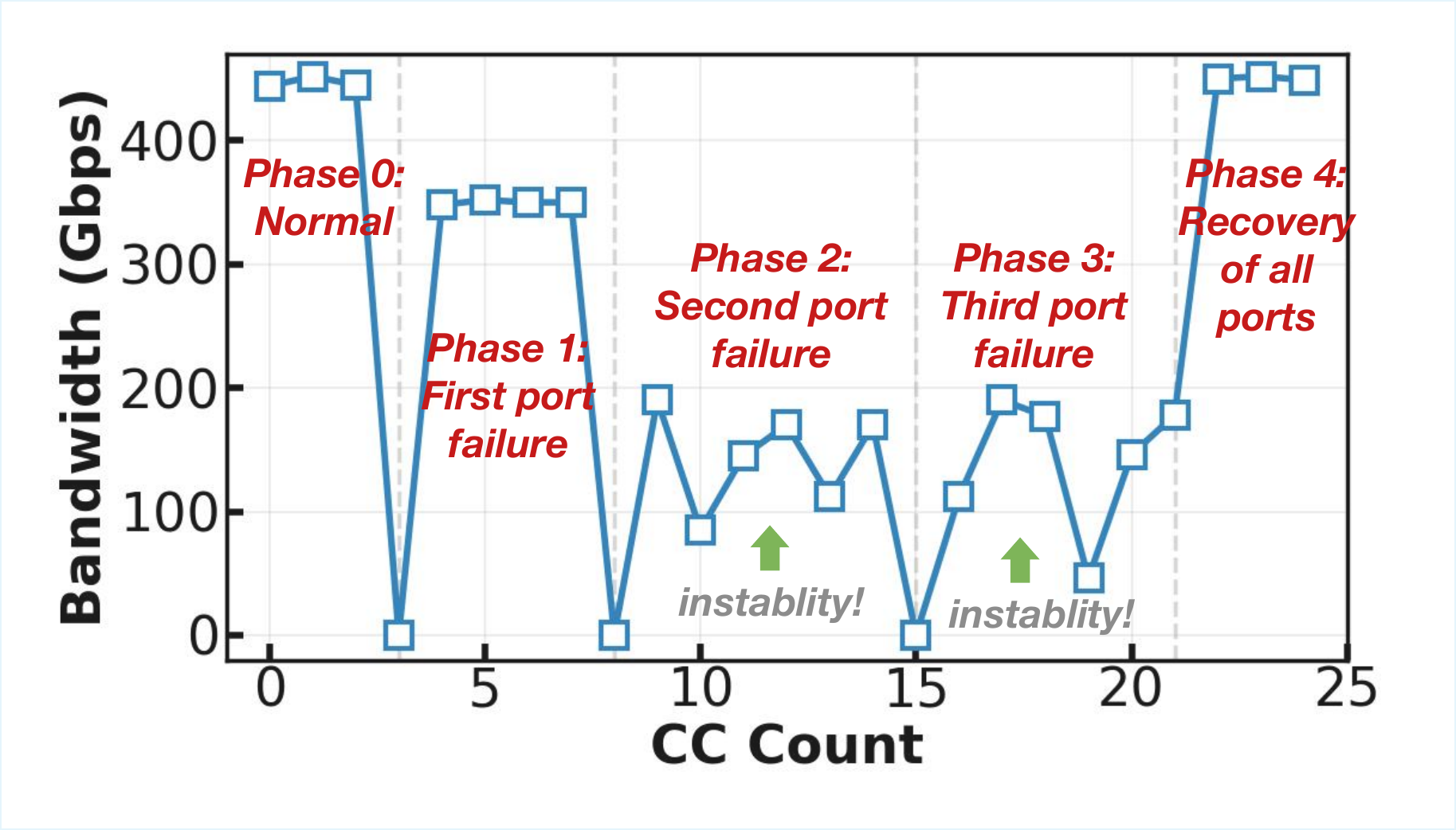}
    \centering
    \vspace{-1\baselineskip}
    \caption{Tolerance to multi-port failures.}
    \label{fig-multi-port-failures}
    \vspace{-1\baselineskip}
\end{figure}

Meanwhile, to evaluate the robustness of our primary-backup mechanism under multi-port failures, we conduct an \texttt{AllRe\allowbreak duce} stress test between two nodes by progressively disabling RNIC ports.
Figure \ref{fig-multi-port-failures} illustrates the bandwidth dynamics across different failure scenarios.
In Phase 0, the system maintains a stable baseline bandwidth of 450 Gbps.
Upon a single port failure (RNIC 0) in phase 1, traffic successfully migrates to the backup QP on RNIC 1, with bandwidth stabilizing at 350 Gbps.
This drop primarily results from physical port sharing and increased PCIe contention as RNIC 1 services two GPUs.
Subsequently, disabling a second port (RNIC 2) in phase 2 triggers a performance degradation to 190 Gbps, characterized by severe instability and jitter.
After careful investigation, we conclude this to congestion collapse, where the asymmetric topology induces many-to-one incast, triggering frequent Priority Flow Control (PFC) \cite{pfc} frames and RDMA backpressure.
Notably, the disabling of the third port (RNIC 4) causes no further performance drop.
This indicates that the bottleneck has shifted from available link capacity to network-wide congestion. 
At this stage, the remaining active paths are fully saturated by the overhead of managing congestion, preventing any further throughput degradation despite fewer available links.
Finally, bringing up all failed ports in phase 4 allows the bandwidth to recover to 450 Gbps, demonstrating that \X can effectively tolerate and recover from multi-port link failures.

\section{Wisdom of Sliding window size}
\label{appendix-sliding-window}

\begin{figure}[h]
    \includegraphics[width=0.99\linewidth]{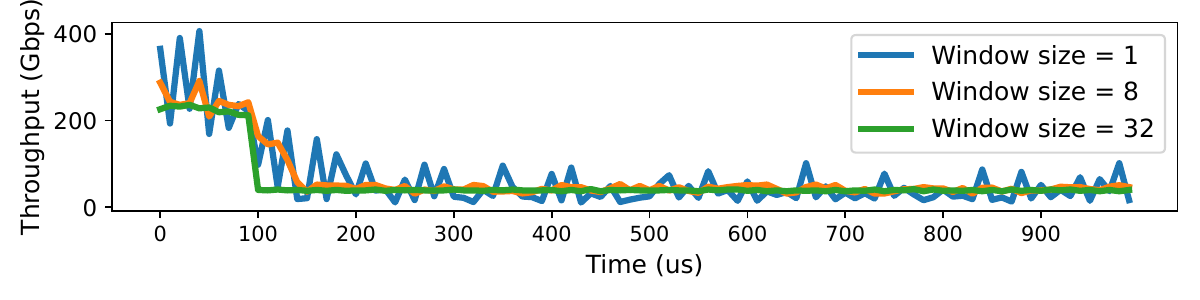}
    \centering
    \vspace{-1\baselineskip}
    \caption{Monitor results with different window sizes.}
    \label{fig-telemetry-throughput}
    \vspace{-1.2\baselineskip}
\end{figure}

To evaluate the observability, we generate a background P2P workload between a pair of servers, and insert a disturbance traffic to compete for bandwidth, making the background workload converge to a lower and stable throughput.
Figure \ref{fig-telemetry-throughput} presents the runtime throughput in a 10$\mu$s granularity with different window sizes.
Especially, when window size is 1, it equals to the naive per-message scheme.
We notice \X can capture the transient $O(\mu s)$ throughput changes, but shows a very fluctuating measuring result with per-message scheme.
With a large window size of 32, \X can smooth out the fluctuation and reflects more accurate results closer to the ground truth, but fails to perceive the instantaneous changes.
For example, when disturbance traffic arrives at 100$\mu$s, per-message scheme shows an intense fluctuation, but a large window size directly drops to the new converged throughput.
Therefore, we set the window size as 8 to achieve the tradeoff between measuring accuracy and fluctuation sensitivity in our environment.

\section{Full-stack Troubleshooting Platform}
\label{appendix-full-stack-troubleshooting}

\begin{figure}[h]
    \centering
    \includegraphics[width=0.98\linewidth]{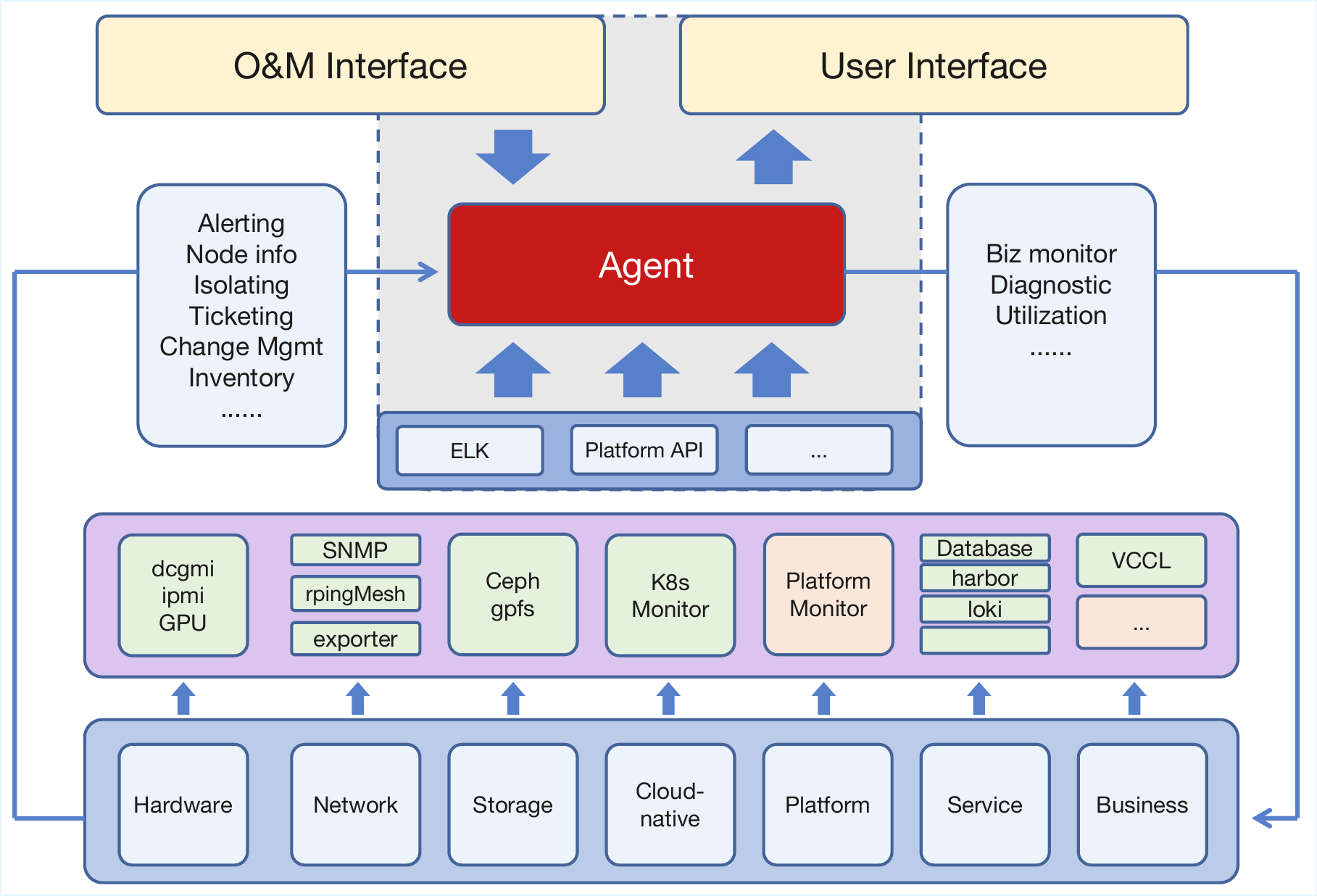}
    \vspace{-0.8\baselineskip}
    \caption{Full-stack trouble shooting platform.}
    \label{fig:trouble_shooting}
\end{figure}

Figure \ref{fig:trouble_shooting} presents the system architecture, organized into three hierarchical layers:
1) Object Layer: covers both physical and logical resources, from hardware (e.g., GPUs, network, storage) to cloud platforms and applications, enabling end-to-end observability.
2) Monitoring and Acquisition Layer (Middle): bridges resources and decisions via specialized collectors.
Standard tools (e.g., dcgmi \cite{dcgmi}, Prometheus\cite{prometheus}) handle general metrics, while \X provides NIC-level troubleshooting and $O(\mu s)$ monitoring capabilities, capturing transient network jitters that are invisible to coarse-grained approaches (e.g., SNMP \cite{SNMP}).
\X also supports customizable application-level metrics (e.g., \texttt{opFuncTimes}, \texttt{ccType}).
3) Decision and Interaction Layer (Top): an intelligent agent aggregates multi-dimensional data via Elastic stacks \cite{elastic_stacks} and platform APIs, driving visualization, alerting, ticketing, and automated fault isolation.

\section{Effectiveness of memory optimization}
\label{appendix-memory-optimization}

\begin{figure}[h]
    \includegraphics[width=0.95\linewidth]{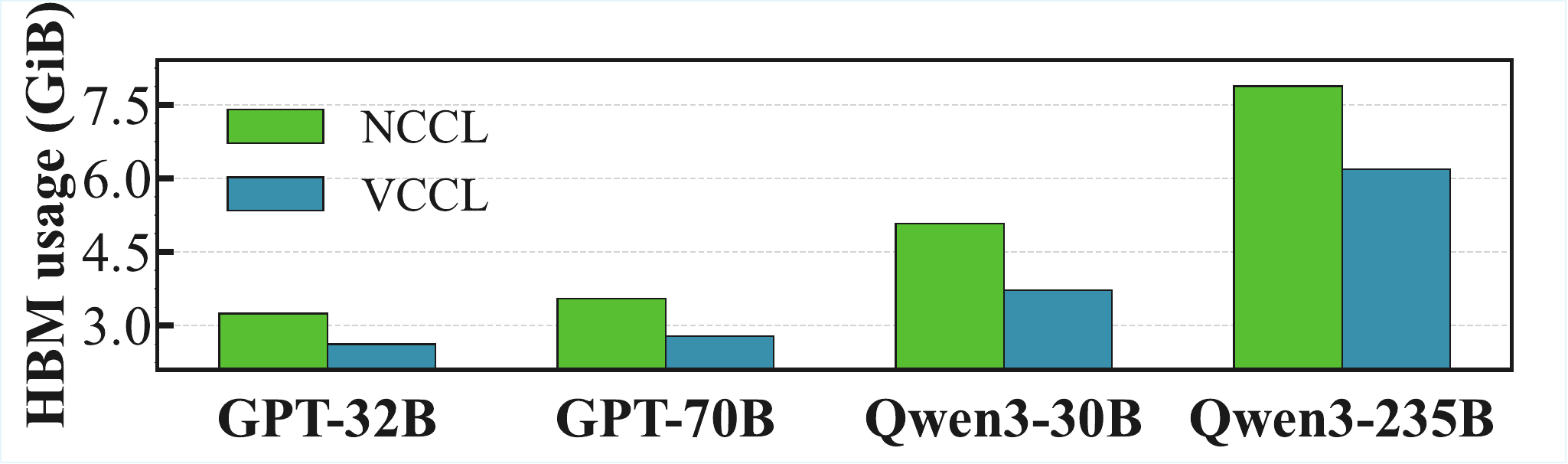}
    \centering
    \vspace{-1\baselineskip}
    \caption{Memory optimization.}
    \label{fig-memory-optimization}
    \vspace{-1\baselineskip}
\end{figure}

To evaluate the effectiveness of the dynamic memory pool in \X, we systematically monitor the memory footprint of both NCCL and \X under a wide range of configurations. 
In particular, we consider two representative dense models, GPT-2 32B and 70B, as well as two MoE models, Qwen3-30B-A3B and 235B-A22B.
As shown in Figure \ref{fig-memory-optimization}, \X achieves up to 26.7\% memory reduction compared to NCCL. 
This result highlights \X’s strong capability in significantly reducing HBM consumption, thereby enabling users to allocate more resources to model execution and focus on higher-level application logic rather than communication-induced memory overhead.

\section{Additional Experience}
\label{appendix-additional-experience}

\noindent\textbf{Proactive isolation is necessary.}
To mitigate the impact of impending hardware failures, we adopt a proactive isolation policy driven by early warning signals. 
For RNIC, we observe that transient link instabilities often precede hard failures. 
Specifically, if an RNIC exhibits two minor flap events within a five-minute window, we immediately mark the host as unhealthy and proactively isolate it. 
The isolation is enforced after the next checkpoint completes, at which point the faulty network interface is replaced. 
This approach allows us to avoid disruptive failures while preserving application progress.
For GPUs, we distinguish between failure modes with different risk profiles. 
Hardware integrity violations, such as ECC errors, trigger immediate isolation due to their strong correlation with irreversible failures and data corruption. 
In contrast, performance degradation caused by environmental factors—such as fan anomalies or transient thermal issues—is handled more conservatively. 
As long as the number of such degradation events does not exceed five occurrences within a 24-hour period, we do not isolate the GPU, allowing the system to tolerate benign fluctuations without unnecessary disruption.

\noindent\textbf{Congestion control.}
As most CC primitives do not incur severe network congestion, (e.g., ring-based \texttt{allreduce}, \texttt{allgather}, \texttt{reducescatter}), we draw some lessons about how to cooperate \X with congestion control, for example, DCQCN \cite{zhu2015congestion}.
On the one hand, when our customers are training dense models that incur mild incasts, tuning DCQCN parameters does not show an obvious improvement on training performance.
As our network architecture already provides sufficient bidirectional bandwidth without oversubscription, we can disable the congestion control and only enable PFC \cite{pfc} to ensure a lossless RoCEv2 environment, like the other .
On the other hand, when our customers are training MoE models with significant \texttt{alltoall} workloads that generate severe incasts, it becomes necessary to tune performant DCQCN parameters for better CC performance.
As a feasible solution, we can drive the DCQCN parameters in throughput-friendly or delay-friendly directions by flow distribution at runtime \cite{chen2024paraleon}.